\documentclass{emulateapj}
\usepackage{apjfonts}
\slugcomment{{\sc To appear in ApJ, 2008 January 1}} 

\newcommand\kms{km~s$^{-1}$}

\newcommand\etal{{et~al.}} 
\newcommand\mM{\ifmmode(m{-}M)\else$(m{-}M)$\fi}

\newcommand\hst{{\it HST}}
\newcommand\HST{{\it HST}}
\newcommand\lta{\lesssim}
\newcommand\gta{\gtrsim}
\newcommand\iacs{\ifmmode I_{814}\else$I_{814}$\fi}
\newcommand\vacs{\ifmmode V_{606}\else$V_{606}$\fi}
\newcommand{\vi}{\ifmmode (V{-}I)\else$(V{-}I)$\fi}
\newcommand{\viz}{\ifmmode (V{-}I)_0\else$(V{-}I)_0$\fi}
\newcommand{\viacs}{(F606W$\,-\,$F814W)}
\newcommand\Mbar{\ifmmode \overline M \else$\overline M$\fi}
\newcommand\MVbar{\ifmmode \overline M_V \else$\overline M_V$\fi}
\newcommand\MIbar{\ifmmode \overline M_I \else$\overline M_I$\fi}
\newcommand\Mvbar{\ifmmode \overline M_V \else$\overline M_V$\fi}
\newcommand\Mibar{\ifmmode \overline M_I \else$\overline M_I$\fi}
\newcommand\ngc{\ifmmode N_{\rm GC} \else$N_{\rm GC}$\fi}

\newcommand\mbar{\ifmmode\overline m\else$\overline m$\fi}
\newcommand\mibar{\ifmmode\overline m_I\else$\overline m_I$\fi}
\def\txitxo{Ben\'{\i}tez}
\newcommand{\Hline}[1]{\mbox{H{\footnotesize {#1}}}}
\newcommand{\Halpha}{\Hline{\mbox{$\alpha$}}}
\newcommand{\HI}{{\sc H\,i}}
\newcommand{\HII}{{\sc H\,ii}}
\newcommand\reff{\ifmmode R_{\rm eff}\else$R_{\rm eff}$\fi}
\newcommand\rhalf{\ifmmode r_h\else$r_h$\fi}

\shortauthors{Barber DeGraaff,~ Blakeslee,~ Meurer~ \&~ Putman}
\shorttitle{NGC 1533: A Galaxy in Transition}

\begin{document}

\title{A Galaxy in Transition: Structure, Globular Clusters, and Distance of the \\ Star-Forming S0 Galaxy NGC 1533 in Dorado\altaffilmark{1}}

\author{Regina Barber DeGraaff\altaffilmark{2},
John P.~Blakeslee\altaffilmark{2},
Gerhardt R.~Meurer\altaffilmark{3},
and
Mary E.~Putman\altaffilmark{4}
}

\altaffiltext{1}{Based on observations made with the NASA/ESA Hubble
Space Telescope, obtained from the Space Telescope Science Institute,
which is operated by the Association of Universities for Research in
Astronomy, Inc., under NASA contract NAS\,5-26555.
These observations are associated with program \#10438.}
\altaffiltext{2}{Dept.\ of Physics \& Astronomy, Washington State University, Pullman, WA 99164; jblakes@wsu.edu}
\altaffiltext{3}{Dept.\ of Physics \& Astronomy, Johns Hopkins University, Baltimore, MD 21218}
\altaffiltext{4}{Department of Astronomy, University of Michigan, Ann Arbor, MI 48109}

\begin{abstract}
We use two-band imaging data from the Advanced Camera for Surveys on board the Hubble
Space Telescope for a detailed study of NGC\,1533, an SB0 galaxy in the Dorado group 
surrounded by a ring of \HI.  NGC\,1533 appears to be completing a
transition from late to early type: it is red, but not quite dead.  Faint spiral
structure becomes visible following galaxy subtraction, and luminous blue stars
can be seen in isolated areas of the disk.  Dust is visible in the color map in
the region around the bar, and there is a linear color gradient throughout the
disk.  We determine an accurate distance from the surface brightness
fluctuations (SBF) method, finding $\mM=31.44\pm0.12$ mag, or $d = 19.4\pm1.1$
Mpc.
We then study the globular cluster (GC) colors, sizes, and luminosity function
(GCLF).  Estimates of the distance from the median of the GC half-light radii
and from the peak of the GCLF both agree well with the SBF distance.  The GC
specific frequency is $S_N=1.3\pm0.2$, typical for an early-type disk galaxy. 
The color
distribution is bimodal, as commonly observed for bright galaxies.
There is a suggestion of the redder GCs having smaller sizes, but the trend is
not significant.  The sizes do increase significantly with galactocentric
radius, in a manner more similar to the Milky Way GC system than to those in
Virgo.  This difference may be an effect of the steeper density gradients in
loose groups as compared to galaxy clusters.
Additional studies of early-type galaxies in low density regions
can help determine if this is indeed a general environmental trend.
\end{abstract}

\section{Introduction}

The Hubble Space Telescope (\hst) has opened the door to our
understanding of extragalactic star cluster systems, revealing numerous
globular clusters (GCs) in early-type galaxies (e.g.\ Gebhardt \&\
Kissler-Patig, 1999; Peng \etal\ 2006) as well as ``super-star clusters''
(SSCs) in late-type galaxies (Larsen \&\ Richtler 2000), especially
starbursts (e.g. Meurer \etal\ 1995, Maoz \etal\ 1996).  In early-type
systems the color distribution of the GCs is often bimodal, consisting of
a blue metal-poor component and a red metal-rich component (e.g. West
\etal\ 2004; Peng \etal\ 2006).  This observation gives a hint to the
connection between the early- and late-type systems.  Mergers often have
particularly strong starbursts and rich populations of SSCs, as seen for
example in NGC4038/39 - ``the Antennae'' system (Whitmore \&\ Schweizer
1995; Whitmore \etal\ 1999), and hierarchical merging is one possible
origin for the redder population of GCs in early-type galaxies 
(e.g., Ashman \& Zepf 1998; Beasley \etal\ 2002; Kravtsov \& Gnedin 2005).

Much of the research on GC systems has concentrated on galaxy clusters which
are rich in early-type galaxies (e.g., the ACS Virgo Cluster Survey,
C{\^o}t{\'e} \etal\ 2004; the ACS Fornax Cluster Survey, Jord{\'a}n \etal\
2007).  Early-type galaxies in groups and the field are somewhat less studied,
particularly with \hst\ and the Wide Field Channel (WFC) of its Advanced
Camera for Surveys (ACS).  The relatively wide ($3\farcm4$) field of view of
the ACS WFC combined with its fine pixel sampling make it an exceptional tool
for imaging GCs out to a few tens of Mpc where they have measurable angular
sizes (Jord{\'a}n \etal\ 2005).

Here we report \hst\ ACS/WFC imaging of the SB0 (barred lenticular) galaxy
NGC~1533 in the Dorado group.  This group is in the ``Fornax wall'' (Kilborn
\etal\ 2005) and hence at a similar distance to the Fornax  cluster
(e.g., Tonry \etal\ 2001).  Dorado is interesting in that it is richer than
the Local Group but still dominated by disk
galaxies (its brightest members being the spiral NGC~1566 and the S0 NGC~1553),
and its members have \HI\ masses similar to non-interacting galaxies with the same
morphology (Kilborn \etal\ 2005).  While the apparent crossing time of the group
is only $\sim 13$\%\ of the age of the universe (Firth \etal\ 2006; see also
Ferguson \&\ Sandage 1990), the most recent analyses conclude the group is
unvirialized (Kilborn \etal\ 2005; Firth \etal\ 2006), which may explain the
richness in spirals and \HI.

\begin{figure*}
\plottwo{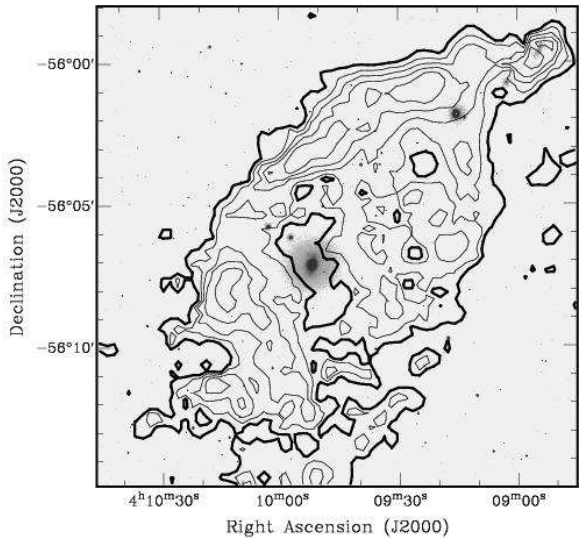}{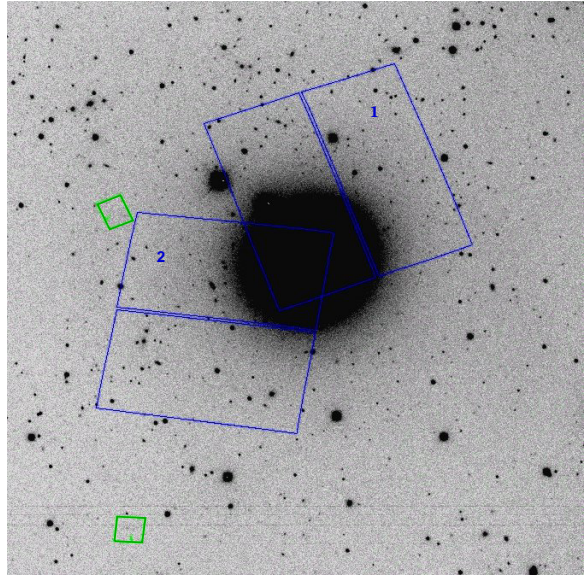}
\caption{\textit{Left panel:} \HI\ contours from Ryan-Weber \etal\ (2003)
are overlaid on a ground-based $R$-band image of the NGC\,1533 field 
from the SINGG survey (Meurer \etal\ 2006).
The outermost contour (bold) is at
a column density of $10^{20}$ cm$^{-2}$, and the contours increase in steps
of $0.5{\times}10^{20}$ cm$^{-2}$.  The small companion galaxies
IC~2038/2039 are in the upper right corner of the image.  NGC\,1533 itself
is in an \HI\ ``hole'' (the galaxy center is not detected),
and this distribution has been described as a ring.
\textit{Right panel:} ACS HRC (green) and WFC (blue) fields of view
for the two HST roll angles described in the text (labeled 1 and 2).
The outlines of the camera fields
are overlaid on a $\sim\,$9\arcmin\ portion of the SINGG $R$-band image.
North is up and East is to the left in both panels.
}
\label{fig:ACSfields}
\end{figure*}

NGC~1533 is the seventh brightest member of the Dorado group,
with $M_V{\,\approx\,}{-}20.7$.
It lies within the virial radius, but is a $\sim\,$2-$\sigma$ velocity
outlier (Kilborn \etal, 2005; Firth \etal\ 2006) so that it is moving
at high speed through the intra-group medium.   A vast \HI\ arc is
seen in the outskirts of NGC~1533 connected to the Sdm galaxy
IC~2038 and the small S0 galaxy IC~2039  (Ryan-Weber \etal\ 2004).
This suggests that NGC~1533 is ``stealing'' ISM from its companions or
has cannibalized another gas-rich satellite
(see Figure~\ref{fig:ACSfields}, left panel).
As is typically seen in S0 galaxies, star formation
is weak in NGC~1533.  Observations of this galaxy in spectroscopic surveys
note the presence of emission lines
(Jorgensen \etal\ 1997; Bernardi \etal\ 2002); the nuclear spectrum
available from the 6dF survey (Jones \etal\ 2005) shows [{\sc N
ii}]6584 and weak \Halpha.  \Halpha\ imaging from the Survey of
Ionization in Neutral Gas Galaxies (SINGG, an \Halpha\ imaging survey of \HI\
selected galaxies; Meurer \etal\ 2006) shows a few weak \HII\ regions
beyond the end of its bar (the nucleus is too bright to allow faint
nuclear \HII\ regions to be detected in the SINGG images), as well as a
scattering of very faint ``intergalactic \HII\ regions''.  These are
discussed in more detail by Ryan-Weber \etal\ (2004) who show that they
are so faint that it would only take one to a few O stars to ionize each one.
Although its current rate is low, the star formation in NGC~1533 illustrates 
another possible channel for building up cluster systems in early-type galaxies:
slow re-ignited star formation in ISM stripped from companions.

The ACS WFC images of NGC~1533 used in the present study were obtained with
\hst\ as ``internal parallel images'' while the ACS High Resolution Channel
(HRC) was pointed at the intergalactic \HII\ regions (\hst\ GO Program 10438;
M. Putman, PI).  The HRC observations are discussed elsewhere 
(Werk \etal\ 2007, in preparation). 
Here we use the WFC observations to measure the structural
properties of the galaxy, characterize its GC population, and use the GC
luminosity function (GCLF), GC half-light radii, and surface brightness
fluctuations (SBF) to provide accurate distance estimates.  The contrast between
NGC~1533, a (weakly) star-forming gas-rich barred S0 in a loose group
environment, and galaxies in the richer environments of the Virgo and Fornax
cluster, provides a useful test of the ubiquity of the various relations
found in the denser environments.

The following section describes the observations and data reductions in more detail.
Sec.~\ref{sec:props} discusses  the galaxy morphology, structure, color profile,
and isophotal parameters.  Sec.~\ref{sec:sbf} presents the SBF analysis and
galaxy distance, while Sec.~\ref{sec:gccolors}--\ref{sec:gclf} discuss 
the GC colors, effective radii, luminosity function, and specific frequency.
The final section summarizes our conclusions.

\begin{figure*}
\epsscale{1.0}
\plotone{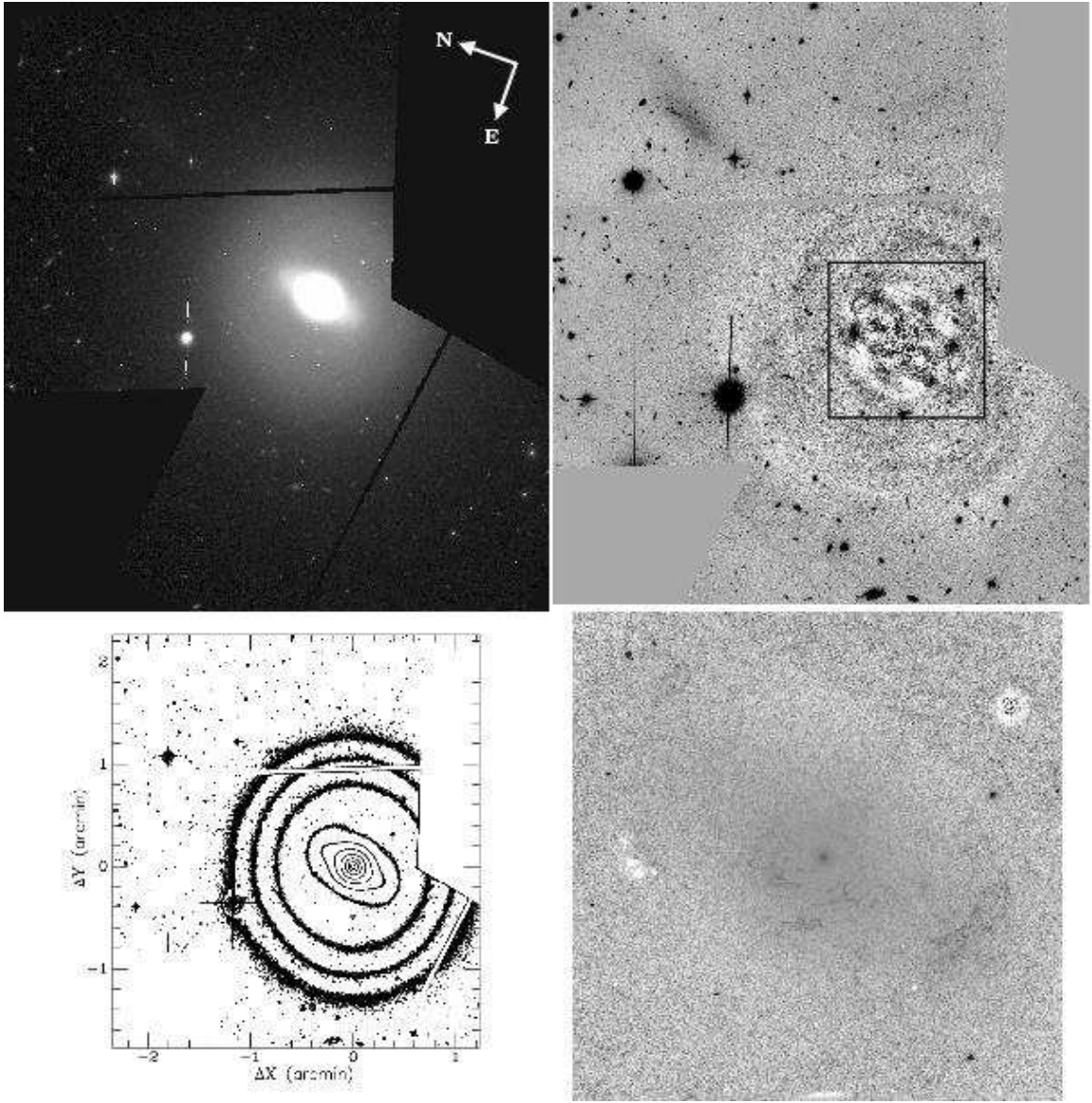}
\caption{\textit{Upper left:} Combined F814W ACS/WFC image of NGC~1533.
% the field size is about 4.8x5.5 arcmin
%
\textit{Lower left:} Contour map of a 3\farcm6$\times$4\farcm0 portion
of the image.  Contours are plotted in steps of a factor of two
in intensity, with the faintest being at $\mu_I = 20.7$ mag~arcsec$^{-2}$.
\textit{Upper right:} The image following galaxy model subtraction,
showing the faint spiral structure (the ``plume''
2\arcmin\ north of the galaxy is a ghost image).  The same 
3\farcm6$\times$4\farcm0 field is shown;
the box marks the central~1\arcmin.
\textit{Lower right:}
Colormap of the central 1\arcmin\ region of NGC1533, with dark
indicating red  areas and white indicating blue. The dark spot
at center marks the center of the galaxy.  Dust can be seen as faint,
dark, wispy features.  A compact blue star-forming region is visible 
to the left of the galaxy center, near the center-left of the map.
}
\label{fig:cmb4}
\end{figure*}
              
\section{Observations and Data Reduction}
\label{sec:obs}

As noted above, during the primary ACS HRC observations of 
\ion{H}{2} regions in the halo of NGC\,1533, the WFC was used for parallel
imaging of the main galaxy.  
These observations were carried out on 2005 September 25 with \HST\
V3-axis position angle PA\_V3 = 110\fdg15 and on 2005 November 18 with PA\_V3 =
170\fdg75; we refer to these throughout as ``roll~1'' and ``roll~2,''
respectively.  Figure~\ref{fig:ACSfields} shows the positions and orientations
of the primary ACS/HRC fields, and the overlapping parallel ACS/WFC fields, at
the two roll angles.

\subsection{Image Processing}

NGC 1533 was imaged in the F814W and F606W bandpasses of the ACS/WFC.  Eight
exposures totaling 4950\,s in F814W, and four exposures totalling 1144\,s in
F606W, were taken at each of the two roll angles.  Following standard
calibration by the STScI archive, the data were processed with the ACS IDT
``Apsis'' pipeline (Blakeslee \etal\ 2003) to produce final, geometrically
corrected, cleaned images with units of accumulated electrons per pixel.
Apsis also ensures the different bandpass images are aligned to better than
0.1~pix and performs automatic astrometric recalibration
of the processed images.

We calibrated the photometry using the Vega-based $m_{1}$ zero points 
from Sirianni et al.\ (2005):  $m_{1,\,F606} = 26.398$
and $m_{1,\,F814} = 25.501$.   
Galactic extinction was taken into account using the dust maps of 
Schlegel \etal\ (1998) and the extinction ratios from Sirianni \etal\ (2005).
We determined extinction corrections of 
$A_{606} = 0.045$ mag and $A_{814} = 0.029$ mag
for the F606W and F814W bandpasses, respectively. 
For comparison to other studies, we also converted the measured
\viacs\ colors to Johnson--Cousins \vi\ using the empirically-based
prescription given by Sirianni et~al.

We processed the data at the two roll angles with Apsis
both separately and combined together.  The upper left panel of
Figure~\ref{fig:cmb4} shows the result from the combined processing,
which is useful for analyzing the galaxy 2-D surface brightness distribution and
isophotal parameters using the largest angular range.  However, for the
SBF and globular cluster analyses, we considered each pointing separately and
then merged the results at the end.  This was done to avoid PSF and orientation
mismatch effects (e.g., diffraction spikes and effects due to the gate
structure of the CCDs do not match up when combining images with differing
orientations).
For each filter image at each pointing, we modeled the galaxy light using the
``elliprof'' software written by J.~Tonry for the SBF Survey of Galaxy
Distances (Tonry \etal\ 1997) and described in more detail by Jord\'an \etal\
(2004).  Saturated areas, bright sources, diffraction spikes, and dusty
regions were masked out for a better model fit.  The galaxy model was then
subtracted from the image, revealing faint sources and residual features,
including faint spiral structure as discussed in Sec.~\ref{sec:props}.

\subsection{Object Photometry}
\label{ssec:objphot}

Object detection was performed with SExtractor (Bertin \& Arnout 1996)
using the galaxy-subtracted image for detection and an RMS image for the weighting.
The RMS image gives the uncertainty per pixel including the effects 
of instrumental and photon shot noise, as well as the additional ``noise''
from the galaxy surface brightness fluctuations.
It is constructed as described in detail by Jord\'an \etal\ (2004):
\begin{equation}
{\rm RMS} \,=\, \sqrt{ ({\rm RMS}_{Ap}) ^{2} + (K_F \ast model) } \,,
\end{equation}
where ${\rm RMS}_{Ap}$ is the Apsis RMS image based
on the instrumental and shot noise alone (Blakeslee \etal\ 2003), \textit{model}
is the galaxy surface brightness model, and $K_F$ gives the ratio
of the variance per pixel from SBF, $\sigma^{2}_{L}$, 
to that from photon shot noise from the galaxy, $\sigma^{2}_{p}$.
The $K_F$ factor depends on the bandpass, exposure time, galaxy distance (which
determines the apparent amplitude of the SBF), and the image resolution;
it can be estimated as
\begin{equation}
K_F \;=\;  \frac{\sigma^{2}_{L}}{\sigma^{2}_{p}}  
    \;=\;  \frac{1}{p}\,10^{-0.4(\overline{m}_F-m^{\ast}_{1,F})} \,,
\label{eq:kf}
\end{equation}
where 
$m^{\ast}_{1,F} = m_{1,F} + 2.5\,\log(T)$, $T$ is the total
exposure time,  $\overline{m}_F$ is the SBF magnitude in the given bandpass,
and $p$ is a factor that reduces the SBF variance because of the
smoothing effect of the PSF.
We adopt $\overline{m}_{F814} \approx 30$ for NGC\,1533 based on the
measurement from Tonry \etal\ (2001) in the very similar $I_C$ bandpass
(and confirmed by our SBF result in Sec~\ref{sec:sbf}) and
assume $\overline{M}_{F606}-\overline{M}_{F814}$ $\approx 2$
based on expectations from stellar population models 
(Liu \etal\ 2000; Blakeslee \etal\ 2001)
to determine $\overline{m}_{F606}$.
Following Jord\'an et al.\ (2004), we convolved simulated noise images with the
ACS PSFs to determine the $p$ factor in Eq.~\ref{eq:kf}; thus, 
we reduced the variance ratio by 12 for F606W and by 13 for F814W
and finally determined the values $K_{F606} = 1.1$ and $K_{F814} = 6.0$.
% $K_{F606W} = 1.095$ and $k_{F606W} = 6.04$.

%%%%%%%%%%%%%
% After some experimentation, 
We ran SExtractor in ``dual image mode,'' such that object detection,
centroiding, and aperture determination was performed only in the deeper
F814W image, while the object photometry was performed in both the
F606W and F814W images.  This ensures that the same pixels are used
for the photometry in both images (see \txitxo\ \etal\ 2004 for a 
detailed discussion).
The catalogs produced by SExtractor include magnitudes measured 
within various fixed and automatic apertures.  By comparing color
measurements of
the same objects at the two different roll angles, we chose to use the
colors measured within an aperture of radius 4~pixels.  Since we are mainly
interested here in compact, only marginally resolved, globular cluster candidates,
we applied a uniform correction of 0.016~mag to the colors measured in this
aperture, based on the difference in the F606W and F814W aperture corrections
(Sirianni \etal\ 2005).  These colors were transformed to \vi\ following
Sirianni \etal, as described above.  For the total $I$-band magnitude,
we use the extinction-corrected, aperture-corrected Vega-based F814W magnitude
\iacs.  According to Sirianni \etal, 
for objects with $\vi\approx1$, this should differ from Cousins
$I_C$ by \hbox{$\lta\,$0.006} mag, or less than the expected zero-point
error of $\sim\,$0.01\,mag.

\begin{figure*}
\epsscale{0.95}
\plotone{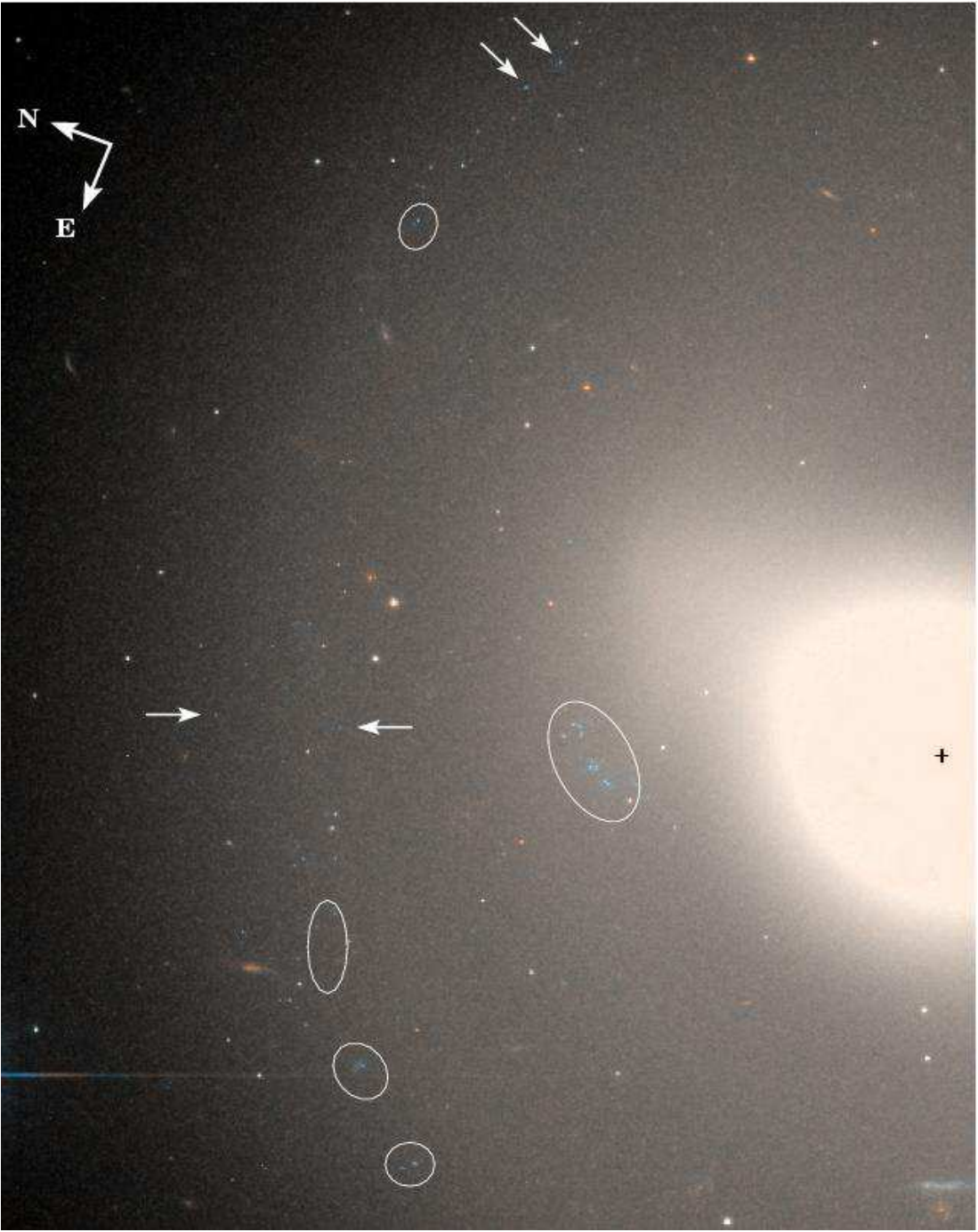}
\caption{\small%
Blue objects in a $1\farcm1{\times}1\farcm4$ region in NGC\,1533.
The ellipses mark the known \ion{H}{2} regions from Meurer \etal\ (2006),
and the arrows point out individual, compact blue objects which appear
to be in the same faint spiral arm. The two white sources within
the largest \HII\ region are GC candidates. The galaxy center is marked and
the orientation is the same as in Fig.~\ref{fig:cmb4}.
}
\label{fig:blueobj}
\end{figure*}

\section{Galaxy Properties}
\label{sec:props}

Tonry \etal\ (2001) reported SBF distances to six members of the Dorado group.
However, one of these (NGC\,1596) was found to be about 20\% closer than the
others; omitting it gives a mean distance modulus for Dorado of $\langle
m{-}M\rangle = 31.40\pm0.09$ mag, or $\langle d\rangle = 19.1\pm0.8$ Mpc.  (We
have revised the published numbers downward by 3\%, as noted in
Sec.~\ref{sec:sbf} below, before taking the average.)  NGC\,1533 itself had a
poorly determined ground-based SBF distance of $d=20.8\pm4.0$, the largest among
the Dorado galaxies, but in agreement with the mean distance given the large
uncertainty.  Based on 24 Dorado group members, Kilborn \etal\ (2005) report a
mean velocity of $1250\pm57$ \kms, with $\sigma = 282$ \kms.  NGC\,1533's
velocity is 790 \kms, indicating a motion through the group of 460~\kms\
towards us.

\subsection{Morphology}\label{ssec:morph}

% say what it's classified as
%   T=-3  (RL)SB0^0

The morphological type of NGC 1533 in the RC3 is $T=-3$ (de Vaucouleurs \etal\
1991), indicating an early-type~S0.  Buta \etal\ (2006) classify it as
(RL)SB0$^0$, meaning that it is a barred, intermediate-type S0 containing
both an inner lens structure and a ring-like feature in the disk.
In a morphological study of 15 early-type disk galaxies,
Laurikainen \etal\ (2006) give an inner radius of 44\arcsec\ for the ring.
Some of these features are evident in the contour map in
Figure~\ref{fig:cmb4}: the round central bulge and convex-lens shape
of its surrounding isophotes, the oblong bar, and the disk are all visible.
In addition, the model-subtracted
image in the upper right panel reveals spiral features that are difficult
to see in the original image.  The faint spiral arms appear to emanate from the
ends of the bar and wrap around by 360$^\circ$.
Sandage \& Brucato (1979) also noted a ``suggestion of weak spiral pattern in outer
lens (or disk)'' in NGC\,1533.  It seems likely that this is
the ``ring'' seen by Buta \etal\ (2006) and Laurikainen \etal\ (2006).
Thus, NGC\,1533 appears to be in the late stages of transition from a barred
spiral to a barred S0 galaxy.  Such morphological transitions are believed to
underlie the observed evolution in the cluster morphology-density relation at
intermediate redshift (e.g., Dressler \etal\ 1997; Postman \etal\ 2005).
Understanding the processes involved in similar transitions at low redshift can
therefore guide our understanding of cluster evolution.

The subtracted image also shows luminous material about 2\farcm0 to the
north/northwest of the NGC\,1533 galaxy center, beyond the galaxy disk.  This
feature is only visible in the F814W image, not in the F606W or ground-based
images.  It is apparently an internal camera reflection from a bright star
outside the field of view.

% describe color map

The color map in the lower right panel of Fig.~\ref{fig:cmb4} reveals the
presence of wispy dust (darker areas in the color map) to the east and
south of the nucleus within the central 20\arcsec\ along the bar.
Moreover, a compact region of bluer light is visible at the center left
edge of the color map, about 23\arcsec\ to the northeast of the
galaxy nucleus.  This position coincides with the brightest \HII\ region
seen in the \Halpha\ image of Meurer \etal\ (2006).  
Our F606W image shows the individual blue stars, as well as
diffuse emission around them in this and the other \HII\ regions.
The diffuse light probably arises from the emission lines 
(especially \Halpha, [{\sc O\,iii}])
in the F606W passband.  The areas of dust and star formation
support the view of NGC\,1533 as a galaxy recently converted from spiral to~S0.
However, it should be noted that dust features are found in roughly
half of bright early-type galaxies when studied at high resolution
(Ferrarese \etal\ 2006), and isolated star formation is not too uncommon.

At larger radius within the disk (outside the color map in the figure), we find
several other small groups, or isolated examples, of unresolved blue objects
with colors $\vi\approx0$ and magnitudes $\iacs\gta23.5$.  Some of these are
visible in Figure~\ref{fig:blueobj}, and the ones associated with \HII\ regions
are marked in the color-magnitude diagram presented in Sec.~\ref{sec:gccolors}.
They may be a mix of post-asymptotic giant branch (PAGB) stars, indicating an
intermediate age population, and small isolated regions of star formation.  
Several of the blue objects, highlighted in Figure~\ref{fig:blueobj},
are spread out along one of the spiral arms.  The likely connection with the
spiral arm in this case points towards these being a small, dispersed group of
fairly young stars.

\begin{figure*}
\epsscale{1.0}
\plottwo{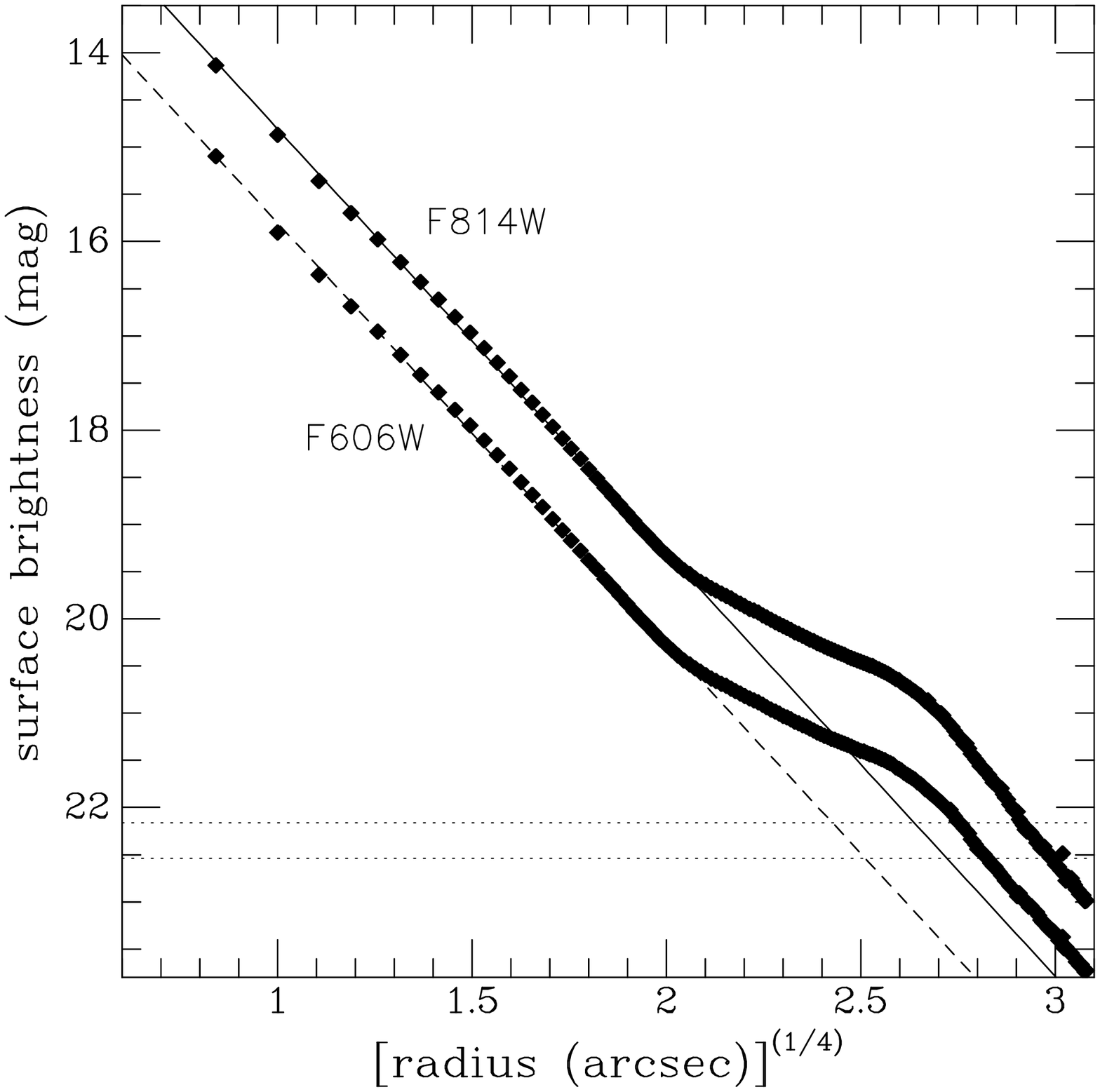}{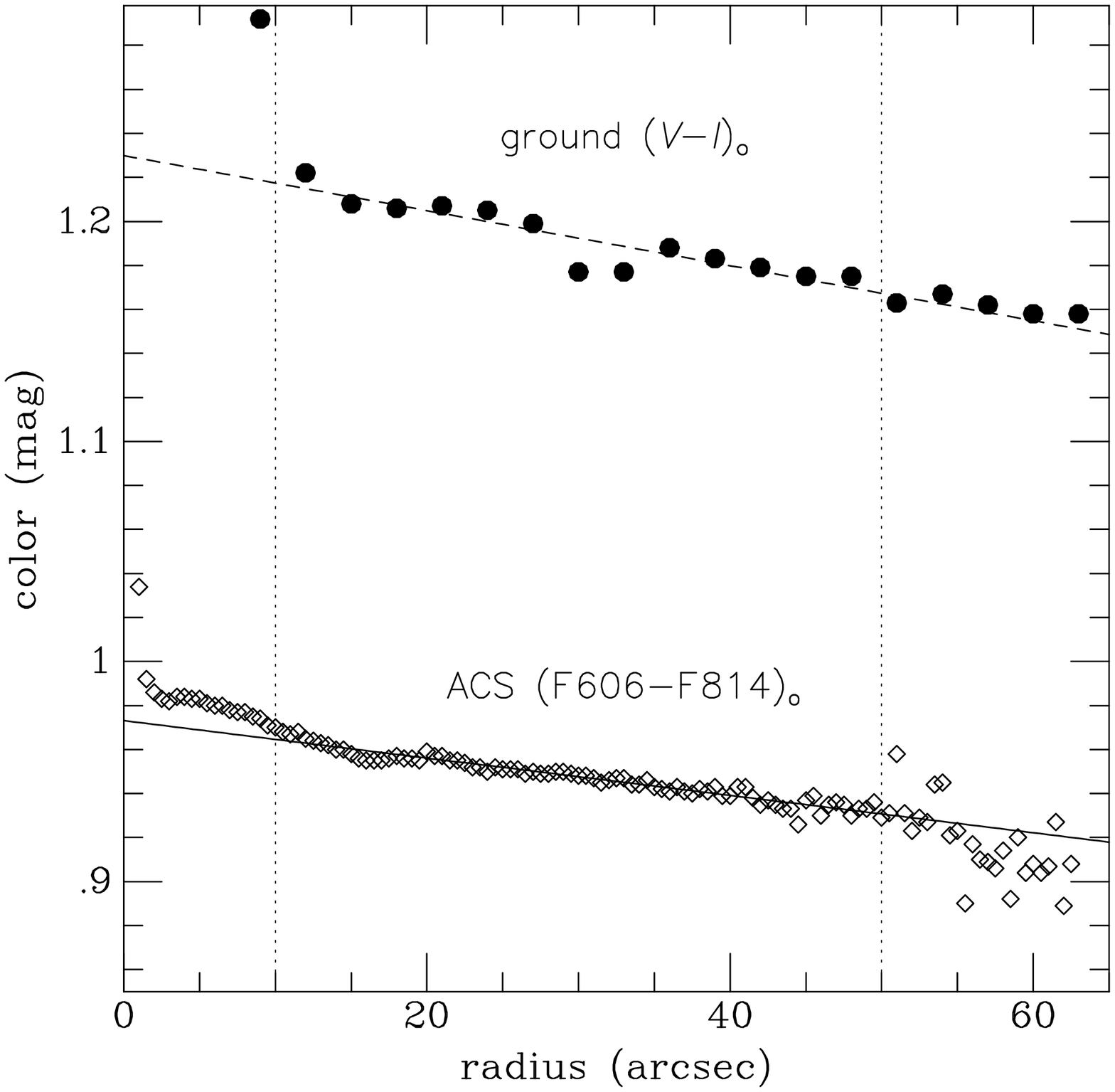}
\caption{
Surface brightnesses in 0\farcs5 circular rings for
the Vega-calibrated F606W and F814W bandpasses are plotted 
as a function of $r^{{1}/{4}}$ (\textit{left}).
The dashed and solid lines
show linear fits to the data inside a radius of 16\arcsec\ where 
the bulge dominates.  The lower and upper dotted lines show
the sky levels in F606W and F814W, respectively.
NGC\,1533 colors from the ground-based \vi\ data (solid circles)
of Tonry \etal\ (1997) and our ACS \viacs\ imaging (open diamonds)
is shown as a function of radius (\textit{right}).
The lines show linear fits to the data points
in the radial range $10\arcsec<r<50\arcsec$.
}
\label{fig:sphot}
\end{figure*}

\begin{figure*}
\epsscale{1.0}
\plottwo{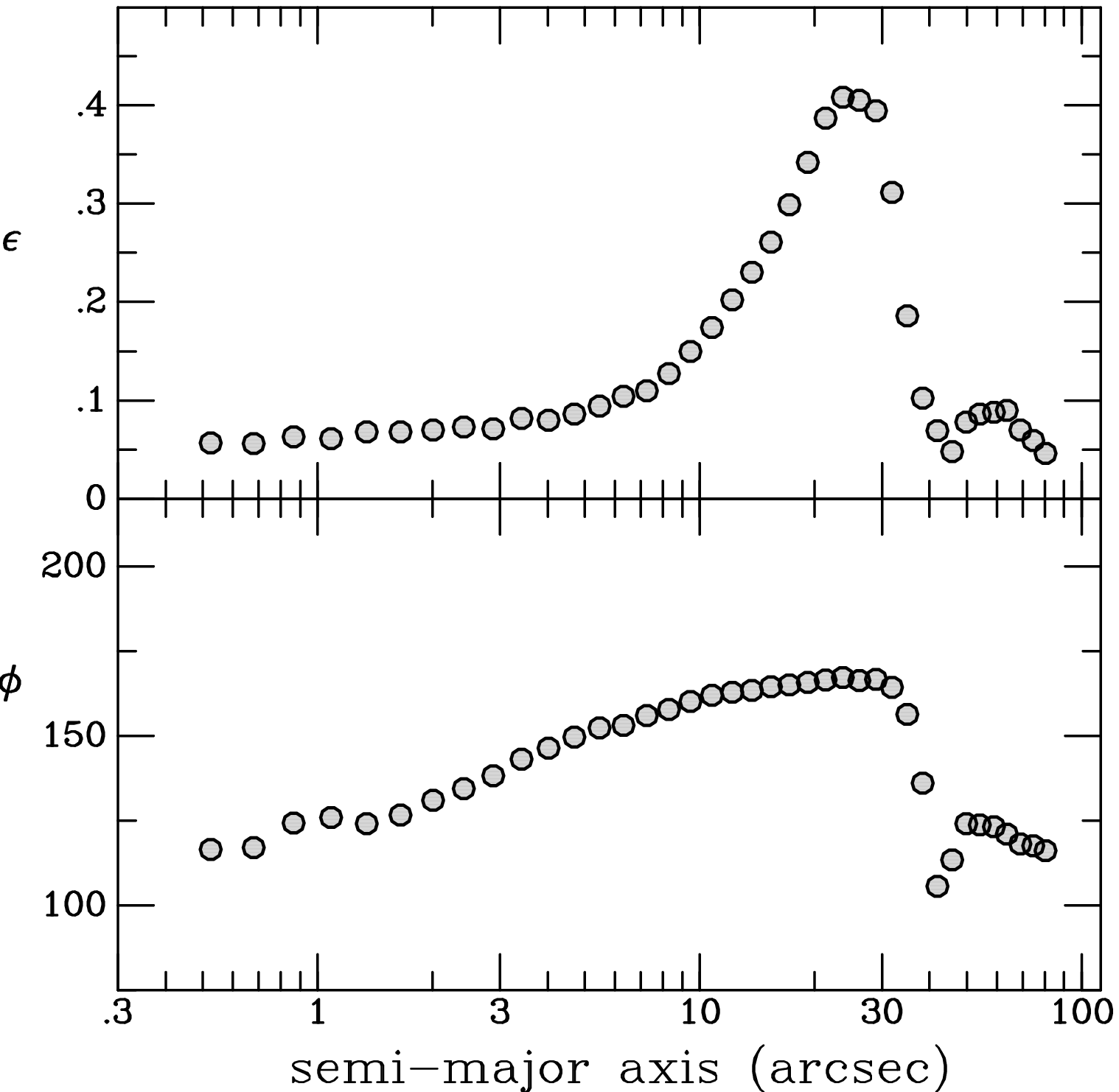}{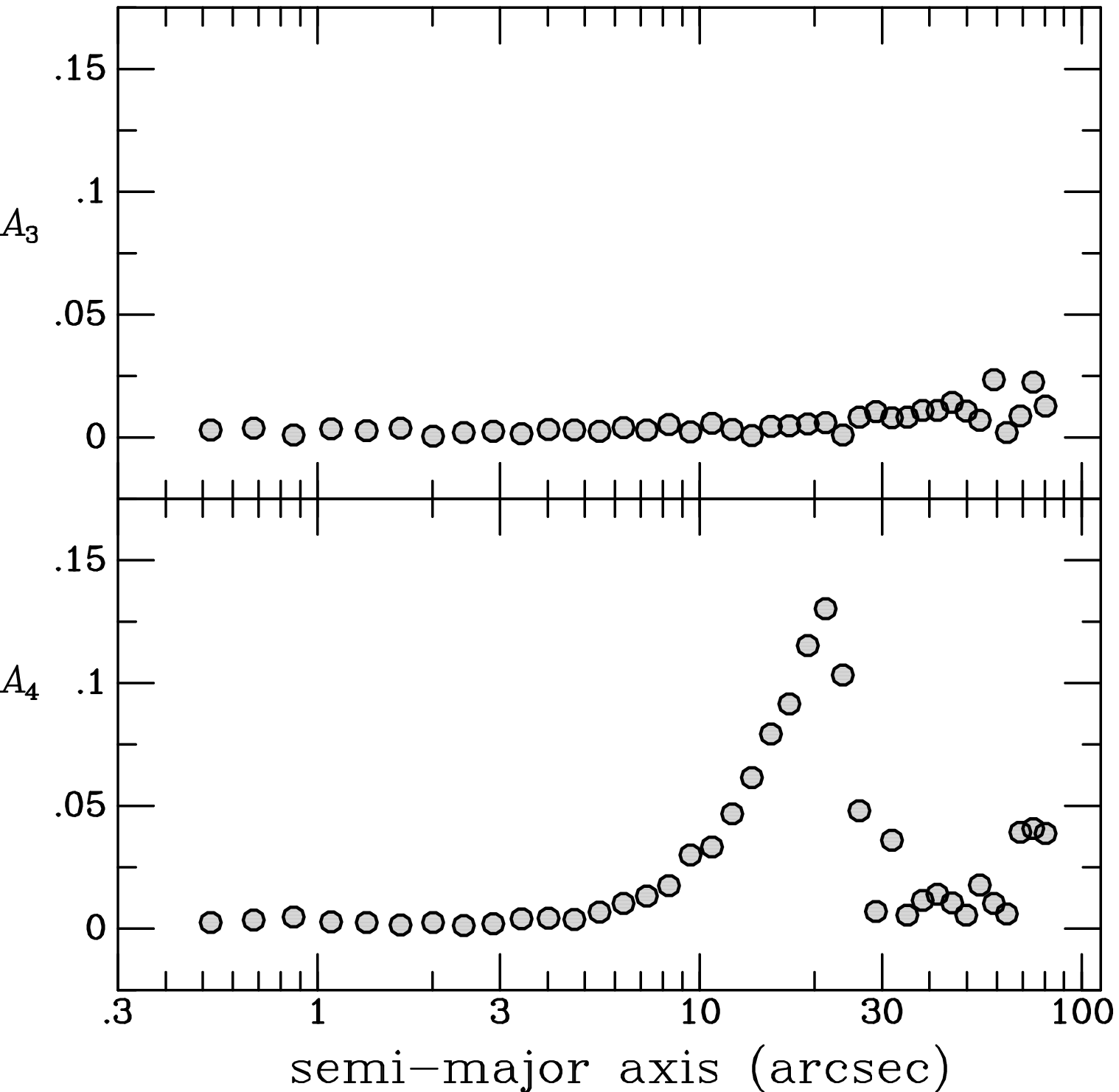}
\caption{Isophotal parameters for NGC 1533.
Ellipticity $\epsilon$ and position angle $\phi$ from the galaxy
isophote modeling are shown versus the semi-major axis of the 
isophote (\textit{left}).
The higher-order $A_3$ and $A_4$ harmonic terms, measuring deviations
of the isophotes from pure ellipses, are shown versus
the isophotal semi-major axis (\textit{left}).
The peak in the $A_4$ profile occurs at 21\arcsec, whereas the
peak in $\epsilon$ occurs at 24\arcsec.
}
\label{fig:isopar}
\end{figure*}

\subsection{Galaxy Surface Photometry and Structure}\label{ssec:colorgrad}
% \subsection{Surface Photometry and Color Gradient}\label{ssec:colorgrad}

Although NGC\,1533 has a complex isophotal structure, it can also be
enlightening to study the simple one-dimensional light profile.
Figure~\ref{fig:sphot} (left panel) shows the galaxy surface brightness profiles
in F606W and F814W.  Within a radius of about 16\arcsec, the profile is
reasonably well fit by a de~Vaucouleurs $r^{1/4}$-law profile (a straight line
in the figure).  Between 16\arcsec\ and $\sim\,$45\arcsec, the profile becomes
much flatter; this includes the inner disk and area around the bar.  In the
outer disk, beyond $\sim\,$50\arcsec, the profile steepens again.

The right panel of Figure~\ref{fig:sphot} shows the galaxy color profile.
In addition to our measured \viacs\ profile, we show the ground-based
\vi\ photometry for this galaxy from the Tonry \etal\ (1997) study (which
provides median and average surface photometry measured in 3\arcsec\ radial bins,
without removal of external sources).
The ground-based data come from multiple long exposures with the Cerro Tololo
4\,m Blanco telescope at a pixel scale of 0\farcs47~pix$^{-1}$.  They suffer
from poor seeing (1\farcs5 in $I$, 1\farcs8 in $V$) and severe central saturation;
however, the systematic error on the color is only 0.018~mag (Tonry \etal\ 2001).
Similar gradients are observed in both sets of photometry.
By binning the data at the same scales and comparing colors over a radial
range of 10\arcsec\,--\,50\arcsec, we determined:
\begin{equation}
\vi \,=\, (1.196\pm0.003) + (1.2\pm0.2)\,[{(\hbox{F606W}-\hbox{F814W})} - 0.95] \,,
\label{eq:vical}
\end{equation}
with an RMS scatter of 0.010~mag in the fit.  This relation yields \vi\ colors
between the empirical (based on stellar photometry) and theoretical 
(based on synthetic spectra and the bandpass transmission curves)
transformations provided by Sirianni \etal\ (2005),
which differ between themselves by $\sim\,$0.07 mag in this color range.
It is in better agreement with the empirical transformation, differing
by about 0.02~mag, as compared to $\sim\,$0.05 mag with the synthetic transformation,
which is probably more uncertain because the F606W bandpass differs substantially
from standard~$V$.
Likewise, Brown \etal\ (2005) reported that when transformations were calculated
based on the bandpass definitions, empirical corrections of order
$\sim\,$0.05 mag were required to match globular cluster data from \vi\ to \viacs.
In any case, Eq.~\ref{eq:vical}
allows for a precise matching of our measured \viacs\ colors to Johnson--Cousins \vi\
over the small range of the galaxy color gradient.  This is important for calibrating
our SBF measurements in Sec.~\ref{sec:sbf} below.

We also performed parametric 2-D surface photometry fits with Galfit (Peng
\etal\ 2002).  For the simplest case, we used a double S\'ersic (1968) model to
fit the bulge and disk (the bar was poorly modeled, mostly as part of the bulge
component).  This analysis yielded a bulge-to-total ratio
$B/T=0.42$, with half-light radii of 7\arcsec\ and 46\arcsec\ for the bulge and
disk, respectively.  The S\'ersic index for the bulge was $n=2.0$, intermediate
between an exponential and a de~Vaucouleurs profile.  However, the fit gave
$n=0.4$ for the disk, or a profile that goes as $\sim \exp(-\alpha r^{2.5})$,
which is even more spatially truncated than a Gaussian.  This agrees with what
was seen in the 1-D plot, where the profile remains fairly constant over a large
radial range then drops off more steeply beyond about 50\arcsec.  We also made
fits with 3 and 4 components.  These gave better model residuals, but the
different components did not neatly break down into clearly distinct physical
components such as bulge, bar, disk, halo (or lens, etc.), so the interpretation
was unclear.

\subsection{Isophotal Parameters}\label{ssec:isopar}

Figure~\ref{fig:isopar} presents the radial profiles of the isophotal
ellipticity, position angle, and $A_3$ and $A_4$ harmonic parameters.
The upper left panel of the figure shows that the galaxy is quite round
within 10\arcsec\ (bulge) and beyond 40\arcsec\ (disk).  However, it
reaches a maximum ellipticity $\epsilon = 0.41$ at a semi-major axis
distance of 24\arcsec, corresponding to the semi-major axis of the bar.
As seen in the lower left panel, there is also a gradual isophotal twist
from $PA \approx 125^\circ$ near a radius of 1\arcsec, to $PA \approx
170^\circ$ where the ellipticity reaches its maximum.  Laurikainen
\etal\ (2006) found similar ellipticity and orientation trends for
NGC\,1533 from their analysis of ground-based $K$-band data, although they
do not appear to have resolved the structure inside a few arcseconds.

The right panels of Figure~\ref{fig:isopar} show the 
amplitudes of the third-order and fourth-order harmonic terms, which
measure the deviations of the isophotes from pure ellipses
(Jedrzejewski 1987).  The values reported by elliprof are the
relative amplitudes of these higher order harmonics with respect
to the mean isophotal intensity, i.e., $A_3 = I_3/I_0$ and
$A_4 = I_4/I_0$. The upper right panel of the figure shows
that the galaxy remains quite symmetric at all radii, since the
$A_3$ component remains near zero.  However, the $A_4$ component,
an indicator of ``diskiness,'' reaches a maximum of 13\% at
a major axis of 21\arcsec, then drops suddenly. 
%
% Thus, the maximum of $A_4$ occurs at smaller radius than
% the maximum ellipticity, indicating that the major axis of
Thus, $A_4$ reaches its maximum at a smaller radius than 
does the ellipticity, since the major axis of
the lens-like isophotes is smaller than that of the bar.  This agrees
with the contour map in Figure~\ref{fig:cmb4}, where the ``convex lens''
shape appears embedded within the oblong bar.

\section{Surface Brightness Fluctuations Distance}
\label{sec:sbf}

We measured the SBF amplitude in four radial annuli for each of the two F814W
observations at the different roll angles.  We used the software and
followed the standard analysis described by Tonry \etal\ (1997), Ajhar \etal\
(1997), Jensen \etal\ (1998), Blakeslee \etal\ (1999, 2001), and references therein. 
More details on the SBF analysis for ACS/WFC data are given
by Mei \etal\ (2005) and Cantiello \etal\ (2005, 2007).
Briefly, after subtracting the galaxy model as described above, we fitted
the large-scale spatial residuals to a two-dimensional grid (we used
the SExtractor sky map for this) and subtracted them to produce a very flat
``residual image.''  All objects above a signal-to-noise threshold
of 10 were removed (masked) from the image.
We used this high threshold to avoid removing the fluctuations themselves,
or the brightest giants in the galaxy. As described
in the following sections, with the
resolution of \HST/ACS we are able to detect and remove globular clusters
(the main source of contamination) to more than a magnitude beyond the peak
of the GCLF, and the residual contamination is negligible.
We also used the F606W image and the color map to identify and mask the dusty
regions and small areas of star formation.

We then modeled the image power spectra in the usual way,
using the template WFC
F814W PSF provided by the ACS IDT (Sirianni \etal\ 2005) and a white
noise component.  We performed the analysis in four radial annuli:
$64<R_p\leq128$,
$128<R_p\leq256$,
$256<R_p\leq512$,
and $512<R_p\leq840$, where $R_p$ is the projected radius in pixels.
The annuli grow by factors of two in order to preserve the same approximate
signal level in each; the outermost limit is set by the proximity of the
galaxy to the edge of the ACS image. 

The SBF amplitude is the ratio of the galaxy image variance (normalization of
the PSF component of the power spectrum) to the surface brightness; it
has units of flux, and is usually converted to a magnitude called \mbar.
The absolute $I$-band SBF magnitude has been carefully calibrated 
according to the galaxy \vi\ color (Tonry \etal\ 1997, 2000).
With a 0.06~mag adjustment to the zero point (Blakeslee \etal\ 2002)
as a consequence of the final revisions in the \hst\ $H_0$ Key Project Cepheid
distances (Freedman \etal\ 2001), the calibration is
\begin{equation}
\Mibar \;=\; -1.68{\pm}0.08 \,+\, (4.5\pm0.25)[\viz - 1.15] \,.
\label{eq:sbfcal}
\end{equation}
For objects with colors similar to the GCs or the mean galaxy 
surface brightness, we have assumed $\iacs = I_C$, since both the 
empirical and synthetic transformations from Sirianni \etal\ (2005)
agree in predicting that the difference should be $<\,$0.01~mag.
However, the SBF is much redder, with a typical color
$\Mvbar{-}\Mibar \approx 2.4$ mag
(Blakeslee \etal\ 2001).  This is outside the color range of the
empirical transformation, but the synthetic one gives
$\Mibar = \overline M_{814} + 0.04$, with an estimated uncertainty 
of $\sim\,$0.02~mag.
We apply this correction and tabulate our SBF results in Table~\ref{tab:sbf}
for the four annuli at the two different roll angles.
The table also gives the galaxy color converted to \vi\
using Eq.~\ref{eq:vical} in the same annuli and with the same 
masking as used for the SBF analysis, and
the resulting distance moduli determined from Eq.~\ref{eq:sbfcal}.

\begin{figure}
\epsscale{0.97}
\plotone{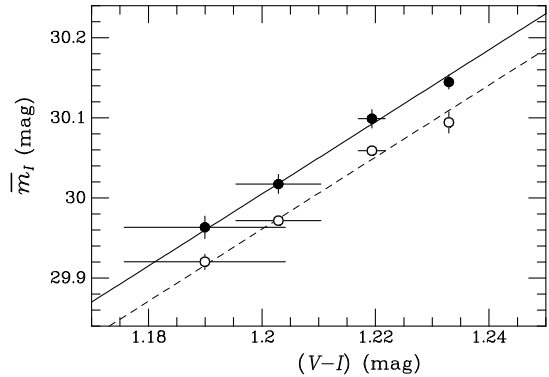}
\caption{SBF measurements in four radial annuli in NGC\,1533
from the roll~1 (filled circles) and roll~2 (open circles) observations.
The solid and dashed lines have slopes of 4.5, as given by the
published \mibar--\vi\ calibration, and are fitted only in the zero~point.
Although internally quite consistent, the two sets of observations
give distance moduli that differ by 0.04~mag.}
\label{fig:sbf}\vspace{-2pt}
\end{figure}

Figure~\ref{fig:sbf} provides a graphical representation the SBF results.
The measurements at the two different roll angles are remarkably
consistent, except for a systematic offset of 0.04~mag.
Weighted averages of the annuli give mean distance moduli and formal
errors of $31.456\pm0.008$ and $31.416\pm0.009$ mag for rolls~1 and~2,
respectively.  The galaxy appears at very different locations in the
field of view for the two observations, in fact on different CCD chips.
We verified that the difference in the photometry itself was negligible
(about ten times smaller than the \mM\ offset).  However, the ACS/WFC
does have some spatial variation in the PSF (Krist 2003), and temporal
variations can be caused by jitter, sun angle, etc.  Any mismatch in
the PSF template used for the power spectrum analysis directly affects
the SBF measurement, and this is the most likely cause of the small
difference.  

We therefore average the results from the two roll angles and use the 0.04~mag
difference as a more realistic estimate of the measurement uncertainty.
To this, we add uncertainties of 0.01~mag in the absolute 
calibration of F814W, 0.02~mag for the transformation
of \mbar\ to the standard $I$~band, 
$4.5\times0.018 = 0.08$ mag from the systematic uncertainty in the \vi\
color used for the calibration (Sec.~\ref{ssec:colorgrad}),
and 0.08 mag from the \Mibar\ calibration zero point.  Finally, we obtain
$(m{-}M) = 31.44 \pm 0.12$ mag, or $d = 19.4 \pm 1.1$ Mpc.
This is an improvement by a factor of 3.5 compared to the ground-based
distance from Tonry \etal\ (2001), and agrees well with the mean SBF
result for the Dorado group (see Sec.~\ref{sec:props} above).  Thus, 
although NGC\,1533 is something of a velocity outlier,
its distance is the same as the group mean.
Our measurement translates to a spatial scale of 94.0\,pc per arcsec,
or 4.70\,pc per ACS/WFC~pixel, which we adopt for the globular cluster
analysis in the following sections.

\section{Globular Cluster Colors}
\label{sec:gccolors}

To obtain a sample of globular cluster candidates from the object photometry
described in Sec.~\ref{ssec:objphot} above, we selected objects with
$19<\iacs<24$ mag, $0.5\leq\vi\leq1.5$, FWHM $\leq$ 4~pix (0\farcs2) in each
bandpass, and galactocentric radius $R_g$ in the range $10\arcsec\ < R_g \leq
108\arcsec$ ($\sim\,$0.9 to $\sim\,$10.2 kpc).  The FWHM selection is about
twice the width of the PSF and is meant to reject clearly extended objects such
as background galaxies.  Correcting for the PSF, it corresponds to an intrinsic
physical FWHM (which is roughly equal to the half-light radius for a typical
King model) of about 16~pc.  This is large enough to include essentially all GCs
(e.g., Jord\'an \etal\ 2005) but may exclude potential ultra-compact dwarfs
(UCDs; Drinkwater \etal\ 2000) such as the ones studied by Ha{\c s}egan \etal\
(2005) in the ACS Virgo Cluster Survey.  A search for UCDs in these data would
be difficult without spectroscopy: they have very low number densities
and there appears to be a distant background galaxy cluster about 2\farcm5
north of NGC\,1533 that would be a source of contamination.

\begin{figure}
\epsscale{0.9}
\plotone{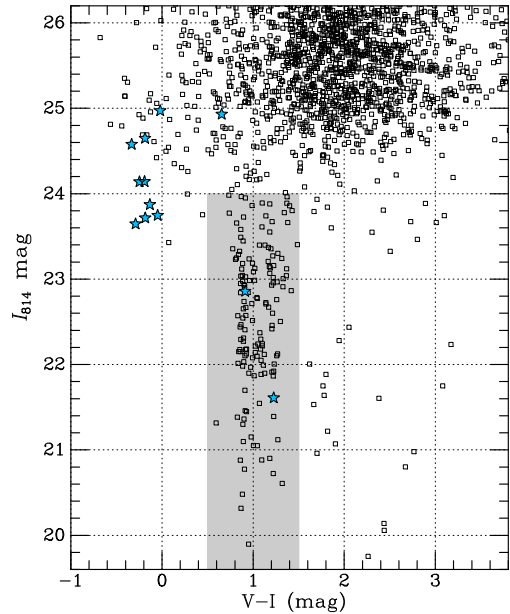}
\caption{Color-magnitude diagram for compact sources (FWHM $<$ 0\farcs2)
in the galactocentric radial range $10\arcsec < R_g < 108\arcsec$.
The shaded region marks the broadest selection we use for GC candidates:
$0.5<\vi<1.5$ and $\iacs<24$.  
The blue stars represent compact sources with $\iacs<25$ 
that lie within the \HII\ regions
identified by Meurer \etal\ (2006).  The two such sources that lie within
the region of the GC candidates are probably simple cases of projection.
In general,
objects with $\vi\approx0\pm0.5$, $\iacs\approx23.5$--25.5 appear to be
blue supergiants either in the \HII\ regions or dispersed along the faint spiral arms;
those with $\vi\approx2$, $\iacs<23$ are likely Galactic M~dwarfs.
The large cloud of points with $\vi\approx2\pm0.1$ and $\iacs\gta25$ 
are (mostly blends of) the brightest evolved giants and supergiants in NGC\,1533.
}\label{fig:bigcmr}
\end{figure}

Figure~\ref{fig:bigcmr} shows the
color--magnitude diagram of sources that have been selected according to
these radial and FWHM constraints.  The shaded region marks the color and
magnitude constraints on the GC candidates. 
The tip of the
red giant branch in NGC\,1533 should occur at $I\approx27.4$, assuming
$M_I^{\rm TRGB}\approx -4$ (Lee \etal\ 1993; Rizzi \etal\ 2007), but
superpositions of multiple red giants, and possibly AGB stars, can be
detected a couple magnitudes brighter than this.  This accounts for the
``cloud'' of faint red sources in Figure~\ref{fig:bigcmr}.

The figure also shows a dozen objects, marked with blue stars, that are
brighter than $\iacs=25$ and lie within the \HII\ regions visible in the
\Halpha\ image of Meurer \etal\ (2006).  These objects have colors $\vi\approx
-0.2\pm0.2$ and absolute magnitudes $M_V\gta-7.5$, which are reasonable for
small associations of a few O and B stars.  Their magnitudes and colors are
similar to the blue stars and associations found in the HRC data
studied by Werk \etal\ (2007, in preparation).
However, the two brightest compact sources found to lie within the \HII\ regions
are actually GC candidates, as shown in the figure.  One of these objects has
the color of a typical ``blue GC'' with $\vi\approx0.9$, while the other is a
typical ``red GC'' with $\vi\approx1.2$.  Both are marginally resolved (i.e.,
nonstellar).  We suspect that this is a simple case of projection and that these
GC candidates are not physically associated with the \HII\ region, since the
projection is within the area of the largest \HII\ region 
(see Figure~\ref{fig:blueobj}), which is located at
a radius where the surface density of GCs is fairly high.
Thus, we treat these two objects the same as the other GC candidates in our
analysis, and simply note that none of our results would change significantly if
they were removed.

\begin{figure}
\epsscale{0.9}
\plotone{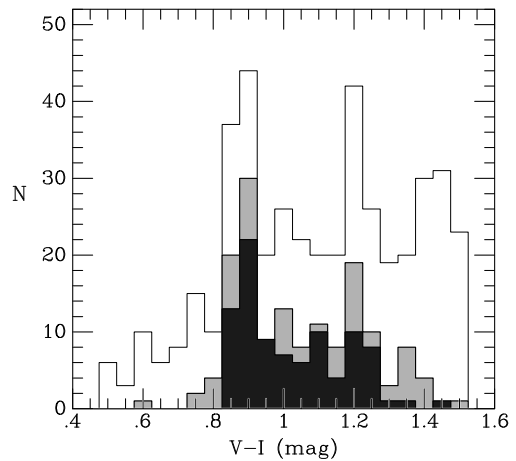}
\caption{Histograms of candidate globular cluster \vi\ colors
(converted from F814W and F606W as described in the text).
The gray histogram uses an $I$-band magnitude cutoff $I<24$
(about 1$\,\sigma$ beyond the peak of the GCLF) with the FWHM and
radial position selections as given in Figure~\ref{fig:bigcmr}.
The open histogram is similar, but uses a cutoff $I<26$, 
and thus has a substantial contribution
from the brightest giants in NGC\,1533.
The dark histogram includes only the $I<23$ GC candidates 
that are found to be nonstellar based on the ISHAPE fits,
and thus should be nearly free from contamination.
}
\label{fig:histo}
\end{figure}

The selection for GC candidates was done for the catalogs from each pointing,
then the two lists of GC candidates were merged, giving a total of 151
candidates.  For objects in the overlapping region of the two roll angles, the
objects' magnitudes and colors were averaged.  Table~\ref{tab:dat} lists the
positions, \vi\ colors, effective radii (discussed in the following section),
\iacs\ magnitudes, and field (roll 1, roll 2, or merge) for each object selected
in this way.
Figure~\ref{fig:histo} shows a histogram of the color distribution of
these objects.  For comparison, it also shows the 
distribution when the magnitude cutoff is $\iacs=26$, and 
the distribution for a cutoff of $\iacs=23$ with apparently stellar sources
(based on the shape fits in Sec.~\ref{ssec:ishape}) removed.
For the former sample ($\iacs{<\,}26$), there is sizable contamination from the 
faint red sources seen in Figure~\ref{fig:bigcmr}, while for the
latter sample ($\iacs{<\,}23$, nonstellar), there should be negligible
contamination.

The \vi\ distribution of the GC candidates (gray histogram in
Figure~\ref{fig:histo}) was tested for bimodality using the KMM
(Kaye's Mixture Modeling) 
algorithm (McLachlan \& Basford 1988; Ashman \etal\ 1994),
which compares the goodness-of-fit between single and double Gaussian
descriptions of the data.  The algorithm finds that the data are better
described by a double Gaussian with $>99.9$\% confidence.  It returns
best-fit values for the color peaks of 0.921 and 1.226~mag. 
Assuming the relation for GCs from Kissler-Patig \etal\ (1998), these correspond
to peak metallicities $[{\rm Fe/H}]\sim-1.5$ and $\sim-0.5$, which are
normal for a galaxy of this luminosity (e.g., Fig.\,13 of Peng \etal\ 2006).
We also considered
restricting the GC candidates to a narrower color range of $0.7 \leq \vi \leq
1.4$ (since the Gaussian assumption makes the method sensitive to outliers) and
to alternative magnitude cutoffs of $\iacs < 23$ and $\iacs < 25$, as well as
excluding the small percentages of likely stellar objects among the candidates.
The confidence of bimodality (in the KMM sense) remained near 100\% for all these
combinations of magnitude and color cuts, except for the case with the faint
magnitude cut $\iacs<25$ and broader color range of $0.5 \leq \vi \leq 1.5$, for which
the confidence level was 98.8\%. This fainter cutoff results in considerable
contamination, yet still exhibits likely bimodality.  We conclude that the GC
color distribution in NGC\,1533 is bimodal to a high level of confidence.

\begin{figure}
\epsscale{0.9}
\plotone{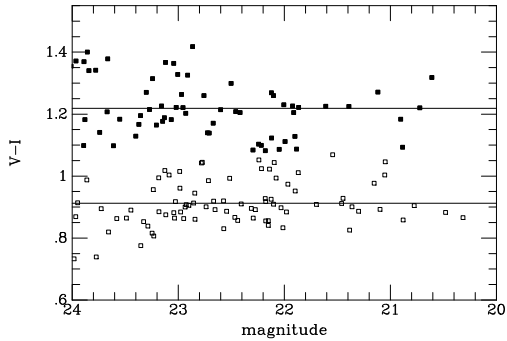}
\caption{\vi\ colors for GC candidates in the red (filled squares)
and blue (open squares) peaks, as determined by the KMM algorithm,
are plotted as a function of $\iacs$ magnitude.
The solid lines show the average colors for the two groups.
There is no significant slope in the color-magnitude relations of
either the red or blue GCs when a cutoff magnitude of $\iacs<23$ is used.
}\label{fig:bluetilt}
\end{figure}

Recently there has been discussion in the literature of a correlation between
magnitude and color for the blue component of the GC population in some
galaxies, in the sense that the blue GCs become redder at higher luminosities
(Harris \etal\ 2006; Strader \etal\ 2006; Mieske \etal\ 2006).  This has been
dubbed ``the blue tilt.''  As a simple test for this in NGC\,1533, we split our
catalog of GC candidates into red and blue groups using the $\vi=1.07$ average
between the two peaks from KMM as the dividing value.  We then performed simple
linear least-squares fits to test for a nonzero slope (see
Figure~\ref{fig:bluetilt}).  With a magnitude cutoff of $\iacs<23$, (just beyond
the peak of the GCLF), we find slopes of $d(V{-}I)/dI = 0.003\pm0.012$ and
$d(V{-}I)/dI = 0.012\pm0.021$ for the color-magnitude relations of the blue and
red GC candidates, respectively; both are zero within the errors.  If we use a
cutoff magnitude of $\iacs<24$, then the slopes both differ from zero by
2$\,\sigma$, but this occurs as a result of the increasing scatter at fainter
magnitudes, coupled with the truncation of the other half the data (simple tests
can reproduce this effect).  When observed, the ``tilt'' occurs because the blue
and red sequences converge at bright magnitudes, not diverge at faint
magnitudes.  There is no evidence of this effect in the present data set.

The lack of the blue tilt for NGC\,1533 is not surprising, since past evidence
for it comes from GCs in the brightest ellipticals in galaxy groups and clusters
(Harris \etal\ 2006; Mieske \etal\ 2006).  In such systems, the GCs reach
higher luminosities and masses, both because the populations are richer
and because the GCLFs are broader (Jord\'an \etal\ 2006).  The simplest
explanation for the tilt is that it results from self-enrichment: the most
massive metal-poor GCs were able to retain some self-enriched gas while star
formation was ongoing.  Since the GCs in a small population like that of
NGC\,1533 do not reach such high masses, this effect may not have occurred.
For instance, Harris \etal\ (2006) report that for very bright galaxies,
the blue and red GC peaks merge into a single broad peak at $M_I<-10.5$. 
In NGC\,1533, only 7~GC candidates have this high luminosity ($I\lta20.9$),
so it impossible to discern if the blue and red peaks merge.
Similarly, if the tilt is due to mergers or accretion by GCs, it might only
occur in the richest systems where these would be more common.  It will be
interesting to see if blue tilts are found in the GC systems of other
intermediate-luminosity galaxies, and if so, whether those systems are
unusually rich or have broad GCLFs.

\section{Globular Cluster Sizes}
\label{sec:gcsize}

\subsection{GC Shape Analysis}
\label{ssec:ishape}

The half-light radius of each GC candidate was obtained using the Ishape program
(Larsen 1999).  Ishape fits the 2-D shape of each object under the assumption
that the object can be modeled by one of various analytic profiles convolved
with the PSF.  We fitted the GC candidates in each roll using the ``KING30''
profile, which is a King (1962) model with concentration parameter $c=30$.
Ishape reports the model FWHM in pixels (prior to PSF convolution), which we
then converted to effective radius \reff\ using the conversion factor of 1.48
given in Table~3 of the Ishape users' manual (Larsen 2005).  We then converted
to a physical size in parsecs using the distance derived in Sec.~\ref{sec:sbf}.
From this analysis, 12 of the 151 GC candidates (7.9\%) brighter than
$\iacs{=\,}24$ were found to be stellar, 
with zero intrinsic size, and were then removed from the catalog.

\begin{figure}
\epsscale{1.0}
\plotone{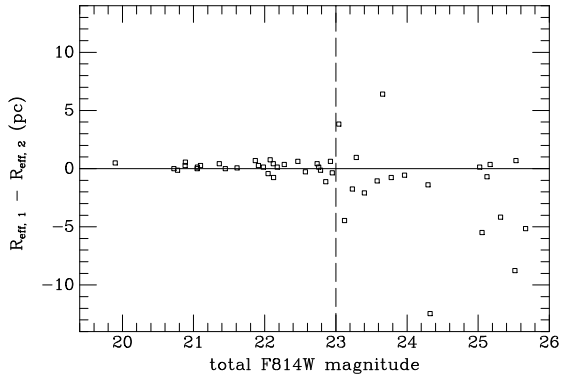}
\caption{Differences in the \reff\ values from Ishape for matched objects present
in both the roll~1 and roll~2 observations are plotted as a function of
\iacs\ magnitude.  There is an apparent abrupt transition from reliable
to dubious measurement values at $\iacs\approx23$.
}\label{fig:Rdiff}
\end{figure}

We fitted the GC candidates (and fainter sources to $\iacs<26$)
for the two different roll angles separately,
then compared the results for sources in the overlapping region.
This is an important test, as Ishape does not provide very robust
size uncertainties.  The manual states that the sizes should be accurate
to about 10\%, given sufficiently high signal-to-noise ($S/N \gta 40$).
Figure~\ref{fig:Rdiff} shows that the agreement is good down to 
$\iacs=23$, where the scatter in the differences is 0.4\,pc (or 0.3\,pc error
per measurement), but worsens abruptly at fainter magnitudes.  The scatter
is larger by a factor of 7.5 for objects with $23<\iacs<24$, and by
a factor of 10 for $23<\iacs<26$.  We therefore consider only objects
with $\iacs<23$ as having reliable \reff\ determinations,
although we tabulate all the measurements in Table~\ref{tab:dat}.

\begin{figure}
\epsscale{0.9}
\plotone{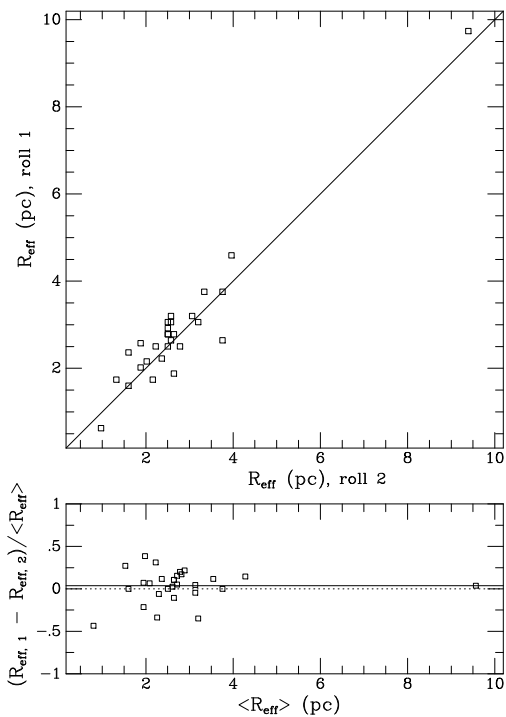}
\caption{\textit{Top:} Measured \reff\ values for objects with $\iacs<23$
in roll~1 are plotted against the \reff\ values for the same
objects measured in roll~2.  The plotted line is equality.
\textit{Bottom:} Fractional differences in the \reff\ values are
plotted as a function of the average value.  The solid line shows
the mean offset of $0.038\pm0.036$.
}\label{fig:split}
\end{figure}

Figure~\ref{fig:split} compares the effective radii from roll~1 to those from
roll~2 for objects with $\iacs<23$ and present in both observations.  The RMS
scatter in the differences is 18.9\%, indicating that the error per measurement
is 13\%.  There is no significant offset in the \reff\ values measured
in the two different observations.

\begin{figure*}
\epsscale{0.72}
\plotone{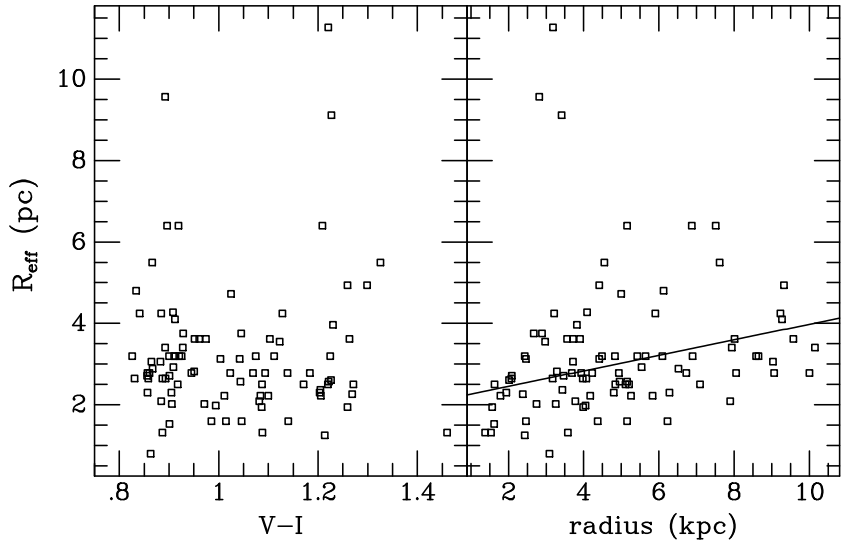}
\caption{\textit{Left:} GC effective radius is plotted versus GC color,
showing no significant correlation.  
\textit{Right:} GC effective radius is plotted versus 
radius from the center of NGC\,1533.  The line shows the correlation
given in Eq.~\ref{eq:reff}, which has a slope of $\sim\,$0.2~pc/kpc.
}\label{fig:sidesplit}
\end{figure*}

Some previous studies have found that $\reff$ depends on the color of the GC,
with red GCs being smaller on average than blue GCs (Jord\'an \etal\ 2005;
Larsen et al 2001).  Figure~\ref{fig:sidesplit} (left panel) shows \reff\ vs
\vi\ for all GCs brighter than $\iacs=23$ in both roll angles (with the values
averaged for objects in the overlap).  We find no statistically significant
trend of \reff\ with color in the present data set.  However, the sample
consists of only 92 objects with robust \reff\ measurements.  If we calculate
the median \reff\ values for the 56 blue GCs and 36 red GCs (using the KMM
splitting from above), we find $\langle \reff \rangle = 2.90\pm0.13$ and
$\langle \reff \rangle = 2.59\pm0.18$ for the blue and red GCs, respectively.
The uncertainties have been estimated by dividing the robust biweight scatter
(Beers \etal\ 1990) by the square root of the number in each sample.  Thus, we
find that the red GCs are $11\pm8\%$ smaller than the blue ones, which is in the
expected sense but not very significant.

On the other hand, we do find a correlation between \reff\ and radial
distance from center of NGC 1533 (Figure~\ref{fig:sidesplit}, right panel).  The
average size of the GCs increases with galactocentric radius.  Omitting the 3
objects at radius $\sim0\farcm5$ with $\reff>9$\,pc (which are more than
6$\,\sigma$ outliers), we find the following relation:
\begin{equation}
\reff \;=\; (2.83\pm0.12) \,+\, (0.191\pm0.047)[(R_g/1\,{\rm kpc}) - 4] \;{\rm pc},
\label{eq:reff}
\end{equation}
where \reff\ is in pc and galactocentric radius $R_g$ is in kpc.
The best-fit slope is virtually unchanged if the blue and red GCs 
are taken separately:
$0.18\pm0.06$ and $0.20\pm0.08$ for blue and red, respectively.
Thus, we find a strong correlation between \reff\ and radius $R_g$,
significant at the  4$\,\sigma$ level, but no significant correlation
between \reff\ and color.  NGC\,1533 is much more similar
to the Milky Way in this regard (e.g., van den Bergh \etal\ 1991) 
than to the early-type galaxies in
the Virgo cluster, where \reff\ has only a mild dependence 
on $R_g$ but a significant dependence on color (Jordan \etal\ 2005). 
It is tempting to associate this difference with environment, but
first it is necessary to study the behavior of \reff\ for the GCs
of many more galaxies in loose group environments.

\subsection{Distance from Half-Light Radius}

Using the extensive ACS Virgo Cluster Survey data set,
Jordan et al.\ (2005) have proposed a distance calibration based
on the median half-light radius of the GC population of a galaxy.
From Eq.~19 of that paper, the distance $d$ in Mpc to the galaxy is estimated as
\begin{equation}
d = \frac{0.552 \pm 0.058}{\langle \hat r_{h}\rangle } \;\hbox{Mpc},
\end{equation}
where $\langle \hat r_{h}\rangle$ is the corrected median half-light radius in arcseconds
(their $r_h$ corresponds to what we have called \reff, following the Ishape notation).
Their definition of $\langle \hat r_{h}\rangle$ involves small
corrections based on galaxy $z$-band surface brightness, galaxy $(g{-}z)$ color,
and GC $(g{-}z)$ color.  We do not have photometry in these bandpasses, and although
we might estimate conversions from models, Jord\'an \etal\ note that the corrections 
are second-order; one can omit them for bright galaxies such as NGC\,1533
and still obtain an accurate distance.
Thus, we simply take $\langle \hat r_{h}\rangle$ as equal to the median
$\langle \reff \rangle = 0\farcs0296\pm0\farcs0011$ 
for the 92 GC candidates in our catalog with $\iacs<23$.
This gives a distance $d  = 18.6 \pm 2.0$ Mpc, in accord with the distance
of $19.4\pm1.1$ Mpc obtained from SBF in Sec.~\ref{sec:sbf}.
The agreement in distance implies that
the sizes of the GCs in NGC\,1533 agree in the median with those in
the Virgo cluster and supports the use of GC half-light radii as distance indicators.

\begin{figure}
\plotone{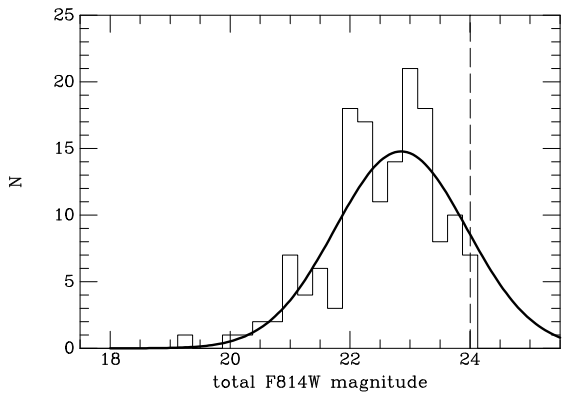}
\caption{The GCLF of candidate globular clusters.
The thick solid curve is a maximum likelihood fit to the
(unbinned) GC magnitude distribution, represented by the
histogram.  The dashed line shows the limiting magnitude
used for the fit.
}\label{fig:GCLF}
\end{figure}

\begin{figure}
\epsscale{1.0}
\plotone{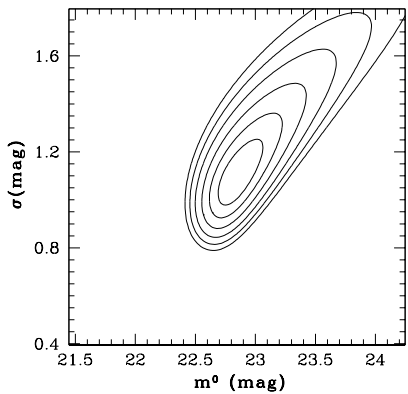}
\vspace{-1.0cm}
\caption{Probability contours on the GCLF width $\sigma$ and turnover
magnitude  $m^0_I$ from the maximum likelihood fitting routine.
The contours are drawn at significance steps of 0.5\,$\sigma$, with the
outermost being at 3$\,\sigma$ (99.7\% confidence).
} 
\label{fig:contours}
\end{figure}

\section{Globular Cluster Luminosity Function}
\label{sec:gclf}

We used the maximum likelihood code from Secker (1992) to fit the
globular cluster luminosity function (GCLF) for 151 GC candidates down
to a limit of $\iacs=24$.  In order to do this, it is necessary to
have a reasonable estimate of the background contamination. We
searched the \HST\ archive for possible background fields with similar
Galactic latitudes ($b\approx-45$) taken through the same F606W+F814W filter
combination to a similar depth.  These fields were processed in the
same way as the NGC\,1533 fields, and the catalogs were subjected to
the same selection according to their magnitude, color, and FWHM.
Some of these fields were found to be anomalously rich, as they
targeted distant rich galaxy clusters; these fields were excluded.  In the
end, we used three high-latitude background comparison fields from
\hst\ program numbers 9405, 9919, and 10438.

For a Gaussian GCLF, we find a turnover (peak) \iacs\ magnitude
$m^0_I = 22.84^{+0.18}_{-0.24}$ and dispersion
$\sigma_{\rm LF} = 1.10 \pm 0.15$ mag.  This GCLF is plotted in
Figure~\ref{fig:GCLF}.
The code also reports the confidence contours on the fit, as shown in
Figure~\ref{fig:contours}.  We performed various tests by 
changing the selection of the data, including narrowing the color
range to be between 0.7 and 1.4, varying the cutoff magnitude by $\pm0.5$ mag,
and being more restrictive with the FWHM cut.  These alternative selections
changed $m^{0}_I$ by about $\pm0.1$ mag, and $\sigma_{\rm LF}$ by about
$\pm0.05$ mag, both well within the quoted errors.

Our SBF distance together with the measured $m^0_I$ implies 
 $M^0_I = -8.6^{+0.22}_{-0.27}$ for NGC\,1533.
Conversely, the GCLF measurement provides 
another estimate of the distance, if we have a calibration
for $M^0_I$.  Harris (2001) gives a $V$-band
calibration $\langle M^0_V\rangle = -7.4\pm0.2$, 
where we use the quoted scatter as an estimate of the uncertainty.
This zero point assumes a Virgo distance modulus
of 30.97, which is 0.12~mag less than the calibration used for the 
ACS Virgo Cluster Survey and in this work.  If we adjust for this
offset and assume $\langle V{-}I\rangle$ = 1.07 mag 
(Sec.~\ref{sec:gccolors}; Gebhardt \& Kissler-Patig 1999), then
we have $\langle M^0_I\rangle = -8.59$, giving $\mM = 31.41\pm{0.29}$~mag. 
This is consistent with the measured SBF distance and the distance 
estimated from the GC half-light radii.  
Note, however, that if we had used
the value of $\langle M^0_V\rangle$ given by Harris for S0 galaxies,
or if we had converted from the value of $\langle M^0_z\rangle$
given by Jord\'an \etal\ (2006), then the inferred distance modulus
would have been larger by about 0.2~mag, although still in agreement
within the errors.

Jord\'an \etal\ (2006) have found a correlation for Virgo galaxies
of the GCLF width with galaxy $B$ luminosity, and we can test whether or
not NGC\,1533 follows this trend. The total apparent $B$-band
magnitude of NGC\,1533 from the RC3 is 11.7. With our measured \mM, Eq.~(2)
from Jord\'an \etal\ (2006) predicts $\sigma_{\rm LF} = 1.09$ mag,
in excellent agreement with our measured value of $1.10\pm0.15$ mag.
We conclude that the GCLF of NGC\,1533 is consistent within the uncertainties
with those observed in Virgo.

Finally, we can estimate the value of the GC specific frequency $S_N = \ngc
\times 10^{0.4(M_V + 15)}$ (Harris \& van den Bergh 1981), where \ngc\ is the
number of globulars and $M_V$ is the absolute magnitude of the galaxy.  The GCLF
analysis indicates that the number of GCs integrated over luminosity is
$163\pm20$ in the region analyzed.  We used an outer radial limit of 1\farcm8
for this study, but a portion of the area within this radius is
missing as a result of the proximity of the galaxy to the image edge at both
roll angles (see Fig.~\ref{fig:ACSfields}) 
and the necessity of omitting the inner 10\arcsec.  If we assume
that the GCs are symmetrically distributed like the galaxy light
(although they may have a more extended density profile at large radii), then 
we estimate a total population $\ngc = 250\pm30$ within 1\farcm8.
The total $V$ magnitude of the galaxy within the same radius is $V =
10.75$, and using the measured distance modulus, we find $S_N = 1.3\pm0.2$.  We
expect this is very close to the ``global'' value, as $\lta\,$5\% of the light
(based on the galaxy profile modeling), and few candidate GCs, are beyond
this radius (which motivated the choice of radius).  This result for NGC\,1533
agrees well with the mean local $S_N = 1.0\pm0.6$ reported by Kundu \& Whitmore
(2001) from \hst/WFPC2 imaging of nearby S0 galaxies and the average
$S_N\approx1.6$ found by Peng \etal\ (2007) 
for intermediate-luminosity early-type (mainly S0) galaxies in Virgo.

\section{Summary and Conclusions}

We have analyzed deep F606W and F814W images of the galaxy NGC\,1533 and its GC
population taken at two roll angles with the ACS/WFC on \hst.  Although it is
classified as an early-type barred lenticular galaxy, we found faint spiral
structure once a smooth fit to the galaxy isophotes was subtracted.  The color
map shows faint dust features in the area around the bar and inner
disk. Previous ground-based \Halpha\ imaging had shown that the galaxy disk
contains several faint, compact \HII\ regions.  We find that all of
these regions have some blue stars associated with them.
Four of the \HII\ regions lie within one of the
faint spiral arms, and several other blue stars are spread out within the arm.  
These observations suggest that NGC\,1533 is in the late stages of a transition 
in morphology from type SBa to~SB0.

Transition objects such as NGC\,1533 may be the key to understanding the
evolution of the morphology-density relation in galaxy clusters, which is often
explained as infalling spirals being transformed into S0s by the harsh cluster
environment (e.g., Dressler \etal\ 1997; Postman \etal\ 2005).  However, recent
evidence at intermediate redshift indicates that the transitions begin outside
the clusters in small group environments through galaxy-galaxy interactions
(Moran \etal\ 2007).  Following infall, the intra-cluster medium then serves to
expedite the process by removing the remaining cool gas.  With NGC\,1533,
we have a close-up view of this transition in the Dorado group, including
interaction with the neighboring galaxies IC~2038/2039 (Ryan-Weber \etal\ 2004).

From two-dimensional two-component parametric modeling of the galaxy surface
brightness, we find a bulge-to-total ratio $B/T\approx0.42$.  The half-light
radii of the bulge and disk are $\sim\,$7\arcsec\ and $\sim\,$46\arcsec,
respectively.  We find a best-fitting S\'ersic index $n=2.0$ for the bulge, 
which can be reasonably approximated by an $r^{1/4}$ law in the 1-D profile.
However, the disk has a relatively flat profile over a factor-of-three in
radius, from $\sim\,$15\arcsec\ to $\sim\,$45\arcsec, then steepens fairly
abruptly beyond $\sim\,$50\arcsec. This gives the disk a very low S\'ersic index
of $n\approx0.4$, which might result from past high-speed interactions 
of NGC\,1533 within the group environment.

Overall, the color of NGC\,1533 is that of an evolved, red population, except in
the few, small isolated regions where the blue stars occur.  The bulge color is
$\vi\gta1.22$, similar to cluster ellipticals, and then there is a mild, but
significant, linear color gradient throughout the disk.  There is a gradual
isophotal twist and the isophotes increase
in ellipticity out to a semi-major axis distance of
24\arcsec, where $\epsilon$ goes above 0.4 before falling sharply again towards
the round outer disk.  The peak of the $A_4$ harmonic term, measuring
``diskiness,'' occurs at a smaller semi-major axis of 21\arcsec.
This is because the pointed lens-like isophotes occur inside of the bar.

We measured the SBF amplitude in four broad radial annuli for each of the two
observations at different roll angles.  A gradient in the SBF amplitude is
clearly detected and follows the color gradient (the bluer outer regions have
relatively brighter SBF).  By matching our ACS photometry against ground-based
\vi\ data for this galaxy, we have accurately calibrated the SBF measurements
to obtain distance moduli.  We find excellent agreement among the different
annuli but with an offset of 0.04 mag in distance between the two observations.
However, the distance error is dominated by systematic uncertainty in
the color and calibration zero point.  We find a final distance
modulus $(m{-}M) = 31.44 \pm 0.12$ mag, or $d = 19.4 \pm 1.1$ Mpc.

Candidate globular clusters were selected according to color, magnitude, radial
position, and FWHM.  Analysis of the color distribution of these objects with
the KMM algorithm indicates with a very high degree of confidence that the
distribution is bimodal with peaks at $\vi\approx0.92$ and 1.22.
There is no evidence that the blue GCs become redder
at bright magnitudes, the so-called ``blue tilt.''  The absence of this effect
in NGC\,1533, an intermediate luminosity galaxy with a small GC population, is
consistent with a self-enrichment explanation, since the GCs in such systems do
not reach the high masses that they do in richer systems.  The sizes of the GC
candidates were measured using the Ishape software. By comparing the results
from the two different roll angles, we found that the effective (half-light)
radii \reff\ have an accuracy of about 13\% down to $\iacs=23$, but are not
reliable beyond this.  We did not find a significant trend of \reff\ with GC
color, although the red-peak GCs have a median \reff\ smaller by $11\pm8$\% than
the blue-peak GCs.

However, we did find a significant (4\,$\sigma$) trend of \reff\ with
galactocentric radius.  In this respect, NGC\,1533 is more like the Milky Way
than the Virgo early-type galaxies.  This may be an effect of the environment:
since the sizes of the GCs are limited by the tidal field, and the density
gradients will be steeper in small groups such as Dorado or the Local Group, GC
sizes should have a stronger dependence on radius in such environments.  The
dominance of this radial effect may weaken or obscure any relation between size
and color.  More studies of size and color trends for the GCs of galaxies in
loose groups are needed to verify this hypothesis, although this may be
difficult because of the low GC populations in such systems.  We then used the
median half-light GC radius to obtain another estimate of the distance to
NGC\,1533.  Following Jord\'an \etal\ (2005), we find $d = 18.6\pm2.0$ Mpc, in
good agreement with the SBF distance.

We modeled the $I_{814}$-band GCLF of NGC\,1533 as a Gaussian using a maximum
likelihood fitting routine. The best-fit peak magnitude 
$m^0_I=22.84^{+0.18}_{-0.24}$ corresponds to $M^0_I \approx -8.6$ for the
measured SBF distance, in good agreement with expectations based on
other galaxies.  The fitted Gaussian dispersion of
$\sigma_{\rm LF} = 1.10 \pm 0.15$ mag is in accord with the relation between
$\sigma_{\rm LF}$ and galaxy luminosity found
recently by Jord\'an \etal\ (2006) for Virgo galaxies.  
Finally we estimate the GC specific frequency in the analysis region to
be $S_N = 1.3\pm0.2$, typical for a disk galaxy.
We conclude that the GCs in NGC\,1533 have the same average size, color,
and luminosity within the errors as the Virgo early-type galaxies, but the
stronger dependence of size on galactocentric distance is more
reminiscent of the Milky Way.  

NGC\,1533 represents an interesting class of transitional objects, both in terms
of morphology and environment.  A large, multi-band, systematic study of such
systems with \hst, similar to the ACS Virgo and Fornax Cluster surveys but
focusing on group galaxies, has yet to be undertaken and must await either a
revived ACS or Wide Field Camera~3.  Such an effort would be extremely
valuable in piecing together a more complete picture of the interplay between
galaxy structure, globular cluster system properties, and environment.

\acknowledgments 
Support for Program
number HST-GO-10438 was provided by NASA through a grant from the Space
Telescope Science Institute which is operated by the Association of
Universities for Research in Astronomy, Incorporated, under NASA
contract NAS5-26555.  GRM acknowledges additional support from NAG5-13083
(the LTSA program).  
We thank Emma Ryan-Weber for providing us with her \HI\ data,
and Andres Jord\'an for helpful comments on the manuscript.

{}

\clearpage

\LongTables

\begin{deluxetable}{rccccc}
%\rotate
%\vspace{-20pt}
\tabletypesize{\small}
\tablewidth{0pt}   % 0pt sets table to its natural width
\tablecaption{SBF Measurements for Various Annuli in NGC 1533}
\tablehead{
\colhead{$\langle r \rangle$\tablenotemark{(a)}} &
\colhead{$(V{-}I)_0$\tablenotemark{(b)}} &
% \colhead{$\pm$} &
\colhead{$\overline m_{I,\,1}$\tablenotemark{(c)}} &
% \colhead{$\pm$} &
\colhead{$\overline m_{I,\,2}$\tablenotemark{(d)}} &
% \colhead{$\pm$} &
\colhead{$(m{-}M)_1$\tablenotemark{(e)}} &
% \colhead{$\pm$} &
\colhead{$(m{-}M)_2$\tablenotemark{(f)}} \\
% \colhead{$\pm$} \\
%
\colhead{(arcsec)} &
\colhead{(mag)} &
\colhead{(mag)} &
\colhead{(mag)} &
\colhead{(mag)} &
\colhead{(mag)} 
}
\startdata
 3.8 & 1.2329 $\pm$ 0.0007 & 30.145 $\pm$ 0.009 & 30.094 $\pm$ 0.013 &  31.452 $\pm$ 0.009 & 31.401 $\pm$ 0.014 \\
 9.2 & 1.2194 $\pm$ 0.0024 & 30.099 $\pm$ 0.011 & 30.059 $\pm$ 0.007 &  31.467 $\pm$ 0.016 & 31.427 $\pm$ 0.012 \\
17.9 & 1.2029 $\pm$ 0.0075 & 30.018 $\pm$ 0.012 & 29.972 $\pm$ 0.006 &  31.460 $\pm$ 0.036 & 31.414 $\pm$ 0.034 \\
32.9 & 1.1899 $\pm$ 0.0142 & 29.963 $\pm$ 0.014 & 29.920 $\pm$ 0.010 &  31.464 $\pm$ 0.065 & 31.421 $\pm$ 0.065 \\
\enddata
\tablecomments{Quoted uncertainties reflect internal measurement error only.
See text for discussion of systematic errors and final averaged distance.}
\tablenotetext{(a)}{Mean radius of annulus.}
\tablenotetext{(b)}{Mean galaxy \viz\ color in annulus.}
\tablenotetext{(c)}{SBF \mibar\ measurement from roll angle~1 observation.}
\tablenotetext{(d)}{SBF \mibar\ measurement from roll angle~2 observation.}
\tablenotetext{(e)}{Distance modulus from galaxy color and roll~1 SBF measurment.}
\tablenotetext{(f)}{Distance modulus from galaxy color and roll~2 SBF measurment.}
\label{tab:sbf}
\end{deluxetable}

%\begin{deluxetable}{crrrrrrrr}\label{tab:data}
\begin{deluxetable}{lcccccccc}
\tablewidth{0pt}
% \tabletypesize{\tiny}
%\tabletypesize{\scriptsize}
\tabletypesize{\small}
\tablecaption{GC Candidates}
\tablehead{
\colhead{ID} & 
\colhead{RA} &
\colhead{DEC} & 
\colhead{$(V{-}I)_0$} & $\pm$ &
\colhead{$R_{eff}$} & 
\colhead{$I_{814}$} &  $\pm$ &
\colhead{field} \\
Number & 
\colhead{(J2000)} &
\colhead{(J2000)} & 
\colhead{(mag)} & (mag) &
\colhead{(arcsec)} & 
\colhead{(mag)} &(mag) 
}
\startdata    
    13   &  62.478742   & $-$56.109037   &    1.260   &    0.042   &   0.0207   &   22.759   &    0.024   &    merge   \\ 
    20   &  62.481901   & $-$56.108821   &    1.341   &    0.079   &   0.0544   &   23.777   &    0.046   &    merge   \\ 
    30   &  62.494871   & $-$56.107931   &    0.867   &    0.028   &   0.0307   &   22.467   &    0.014   &    merge   \\ 
   141   &  62.503620   & $-$56.108303   &    1.082   &    0.025   &   0.0222   &   22.180   &    0.012   &       r2   \\ 
   175   &  62.469642   & $-$56.111536   &    0.913   &    0.052   &   0.0340   &   22.856   &    0.034   &    merge   \\ 
   199   &  62.475854   & $-$56.111165   &    0.863   &    0.049   &   0.0085   &   22.950   &    0.032   &    merge   \\ 
   219   &  62.456702   & $-$56.113055   &    0.863   &    0.088   &   0.0159   &   23.580   &    0.058   &    merge   \\ 
   391   &  62.458113   & $-$56.114010   &    0.855   &    0.037   &   0.0289   &   22.176   &    0.023   &    merge   \\ 
   424   &  62.477575   & $-$56.112507   &    1.123   &    0.034   &   0.0377   &   22.120   &    0.019   &    merge   \\ 
   428   &  62.512621   & $-$56.109436   &    1.299   &    0.025   &   0.0525   &   22.503   &    0.011   &       r2   \\ 
   443   &  62.468860   & $-$56.113378   &    1.012   &    0.036   &   0.0237   &   21.869   &    0.022   &    merge   \\ 
   467   &  62.466586   & $-$56.113788   &    1.086   &    0.046   &   0.0207   &   22.049   &    0.029   &    merge   \\ 
   472   &  62.470944   & $-$56.113139   &    1.226   &    0.033   &   0.0277   &   21.610   &    0.018   &    merge   \\ 
   556   &  62.498873   & $-$56.111763   &    0.816   &    0.036   &   0.0215   &   23.248   &    0.021   &       r2   \\ 
   566   &  62.486962   & $-$56.112939   &    1.276   &    0.084   &   0.0592   &   23.898   &    0.054   &       r2   \\ 
   616   &  62.507189   & $-$56.111401   &    0.987   &    0.045   &   0.0629   &   23.863   &    0.026   &       r2   \\ 
   785   &  62.488994   & $-$56.114282   &    0.806   &    0.080   &   0.0433   &   23.231   &    0.060   &    merge   \\ 
   883   &  62.473735   & $-$56.116447   &    0.901   &    0.050   &   0.0163   &   22.738   &    0.034   &    merge   \\ 
  1030   &  62.487940   & $-$56.116245   &    1.093   &    0.023   &   0.0296   &   20.884   &    0.011   &    merge   \\ 
  1048   &  62.491160   & $-$56.116023   &    0.869   &    0.072   &   0.0355   &   23.966   &    0.052   &    merge   \\ 
  1151   &  62.483266   & $-$56.117620   &    0.950   &    0.024   &   0.0300   &   19.897   &    0.011   &    merge   \\ 
  1238   &  62.494429   & $-$56.117187   &    0.895   &    0.056   &   0.0696   &   23.727   &    0.039   &       r2   \\ 
  1336   &  62.474199   & $-$56.119748   &    1.220   &    0.025   &   0.0266   &   20.723   &    0.013   &    merge   \\ 
  1388   &  62.475712   & $-$56.120119   &    0.905   &    0.026   &   0.0244   &   20.775   &    0.013   &    merge   \\ 
  1398   &  62.487527   & $-$56.119145   &    0.907   &    0.045   &   0.0455   &   22.925   &    0.028   &    merge   \\ 
  1542   &  62.472319   & $-$56.121682   &    1.070   &    0.090   &   0.0355   &   23.399   &    0.065   &    merge   \\ 
  1550   &  62.469259   & $-$56.122013   &    1.088   &    0.054   &   0.0141   &   22.218   &    0.036   &       r2   \\ 
  1607   &  62.483549   & $-$56.121258   &    0.901   &    0.025   &   0.0289   &   21.363   &    0.013   &    merge   \\ 
  1633   &  62.490568   & $-$56.120816   &    1.270   &    0.060   &   0.0037   &   23.301   &    0.033   &       r2   \\ 
  1645   &  62.468214   & $-$56.122742   &    1.460   &    0.085   &   0.0141   &   22.868   &    0.053   &       r2   \\ 
  1669   &  62.486759   & $-$56.121398   &    0.858   &    0.023   &   0.0281   &   20.880   &    0.011   &    merge   \\ 
  1690   &  62.483049   & $-$56.121813   &    1.205   &    0.031   &   0.0252   &   21.911   &    0.017   &    merge   \\ 
  1692   &  62.489449   & $-$56.121244   &    1.326   &    0.045   &   0.0585   &   22.910   &    0.026   &       r2   \\ 
  1760   &  62.490182   & $-$56.121732   &    1.141   &    0.074   &   0.0185   &   23.741   &    0.047   &       r2   \\ 
  1794   &  62.459528   & $-$56.124760   &    1.046   &    0.026   &   0.0170   &   21.050   &    0.013   &    merge   \\ 
  1822   &  62.493954   & $-$56.121881   &    0.900   &    0.038   &   0.0340   &   22.935   &    0.023   &       r2   \\ 
  1837   &  62.482462   & $-$56.122994   &    0.839   &    0.057   &   0.0348   &   23.287   &    0.039   &    merge   \\ 
  1847   &  62.497809   & $-$56.121721   &    0.834   &    0.025   &   0.0511   &   22.013   &    0.013   &       r2   \\ 
  1977   &  62.466685   & $-$56.125414   &    1.269   &    0.041   &   0.0240   &   22.123   &    0.025   &    merge   \\ 
  1980   &  62.475828   & $-$56.124632   &    0.892   &    0.036   &   0.1017   &   22.274   &    0.022   &    merge   \\ 
  2004   &  62.467446   & $-$56.125627   &    1.042   &    0.055   &   0.0333   &   22.783   &    0.036   &    merge   \\ 
  2023   &  62.495436   & $-$56.123193   &    1.183   &    0.054   &   0.0237   &   23.552   &    0.033   &       r2   \\ 
  2032   &  62.472654   & $-$56.125346   &    0.928   &    0.027   &   0.0400   &   21.448   &    0.015   &    merge   \\ 
  2043   &  62.483985   & $-$56.124394   &    0.893   &    0.024   &   0.0281   &   21.098   &    0.012   &    merge   \\ 
  2101   &  62.480199   & $-$56.125032   &    0.882   &    0.050   &   0.1343   &   23.042   &    0.033   &    merge   \\ 
  2138   &  62.460701   & $-$56.127096   &    1.367   &    0.058   &   0.0474   &   23.121   &    0.035   &    merge   \\ 
  2215   &  62.486362   & $-$56.125426   &    0.819   &    0.095   &   0.0585   &   23.657   &    0.045   &    merge   \\ 
  2262   &  62.472163   & $-$56.127100   &    0.831   &    0.041   &   0.0281   &   22.571   &    0.026   &    merge   \\ 
  2321   &  62.496776   & $-$56.125300   &    0.857   &    0.029   &   0.0244   &   22.439   &    0.015   &       r2   \\ 
  2536   &  62.470591   & $-$56.129244   &    0.884   &    0.030   &   0.0222   &   21.981   &    0.017   &    merge   \\ 
  2613   &  62.474279   & $-$56.129425   &    0.994   &    0.033   &   0.0211   &   22.078   &    0.018   &    merge   \\ 
  2657   &  62.484181   & $-$56.128924   &    1.222   &    0.028   &   0.0274   &   21.865   &    0.014   &       r2   \\ 
  2664   &  62.494131   & $-$56.128090   &    1.215   &    0.045   &   0.0562   &   23.272   &    0.025   &       r2   \\ 
  2720   &  62.475693   & $-$56.130236   &    1.004   &    0.172   &   0.0333   &   21.053   &    0.012   &    merge   \\ 
  2753   &  62.515850   & $-$56.126848   &    1.195   &    0.035   &   0.0895   &   23.355   &    0.017   &       r2   \\ 
  2782   &  62.509445   & $-$56.127683   &    1.129   &    0.036   &   0.0385   &   23.399   &    0.019   &       r2   \\ 
  2804   &  62.467737   & $-$56.131626   &    0.826   &    0.024   &   0.0340   &   21.384   &    0.012   &       r2   \\ 
  2846   &  62.510464   & $-$56.128097   &    0.864   &    0.024   &   0.0326   &   22.293   &    0.011   &       r2   \\ 
  2884   &  62.462936   & $-$56.132600   &    0.918   &    0.030   &   0.0266   &   22.174   &    0.016   &       r2   \\ 
  2959   &  62.466327   & $-$56.132967   &    1.023   &    0.030   &   0.0296   &   22.141   &    0.016   &       r2   \\ 
  3108   &  62.470197   & $-$56.133570   &    0.890   &    0.051   &   0.0141   &   23.448   &    0.033   &       r2   \\ 
  3287   &  62.479226   & $-$56.134192   &    0.841   &    0.025   &   0.0451   &   22.151   &    0.013   &       r2   \\ 
  3702   &  62.464106   & $-$56.138694   &    0.896   &    0.025   &   0.0681   &   22.311   &    0.012   &       r2   \\ 
  4233   &  62.482340   & $-$56.141013   &    0.739   &    0.045   &   0.0710   &   23.773   &    0.025   &       r2   \\ 
  4244   &  62.494051   & $-$56.140087   &    0.861   &    0.028   &   0.0296   &   22.841   &    0.014   &       r2   \\ 
  4269   &  62.504504   & $-$56.139280   &    0.875   &    0.031   &   0.0237   &   23.117   &    0.015   &       r2   \\ 
  4737   &  62.473405   & $-$56.145470   &    0.912   &    0.020   &   0.0437   &   21.459   &    0.009   &       r2   \\ 
  4740   &  62.493197   & $-$56.143711   &    1.069   &    0.023   &   0.0296   &   21.544   &    0.009   &       r2   \\ 
  4791   &  62.482879   & $-$56.145036   &    0.961   &    0.032   &   0.0385   &   22.984   &    0.015   &       r2   \\ 
  2072   &  62.413458   & $-$56.114438   &    0.775   &    0.032   &   0.0244   &   23.354   &    0.017   &       r1   \\ 
  2617   &  62.429086   & $-$56.100548   &    1.128   &    0.021   &   0.0451   &   21.899   &    0.009   &       r1   \\ 
  2645   &  62.422291   & $-$56.111412   &    1.225   &    0.020   &   0.0340   &   21.391   &    0.009   &       r1   \\ 
  2859   &  62.433442   & $-$56.097663   &    1.207   &    0.045   &   0.0511   &   23.672   &    0.021   &       r1   \\ 
  3048   &  62.430437   & $-$56.105507   &    1.103   &    0.023   &   0.0385   &   22.242   &    0.011   &       r1   \\ 
  3055   &  62.425954   & $-$56.112310   &    0.956   &    0.034   &   0.0385   &   23.236   &    0.018   &       r1   \\ 
  3110   &  62.427170   & $-$56.111496   &    0.864   &    0.039   &   0.0437   &   23.489   &    0.022   &       r1   \\ 
  3125   &  62.427697   & $-$56.110888   &    1.018   &    0.032   &   0.0222   &   23.128   &    0.018   &       r1   \\ 
  3143   &  62.428683   & $-$56.109900   &    0.866   &    0.017   &   0.0585   &   20.316   &    0.007   &       r1   \\ 
  3540   &  62.430444   & $-$56.113735   &    0.925   &    0.023   &   0.0340   &   22.125   &    0.012   &       r1   \\ 
  3645   &  62.443910   & $-$56.095340   &    0.914   &    0.047   &   0.0511   &   23.949   &    0.025   &       r1   \\ 
  3647   &  62.430762   & $-$56.115511   &    1.139   &    0.030   &   0.0296   &   22.708   &    0.015   &       r1   \\ 
  3978   &  62.440401   & $-$56.106818   &    1.015   &    0.036   &   0.0170   &   22.986   &    0.021   &       r1   \\ 
  4063   &  62.433704   & $-$56.118506   &    0.920   &    0.028   &   0.0340   &   22.574   &    0.016   &       r1   \\ 
  4114   &  62.436845   & $-$56.114726   &    1.074   &    0.029   &   0.0340   &   22.371   &    0.015   &       r1   \\ 
  4202   &  62.454829   & $-$56.089105   &    0.927   &    0.022   &   0.0363   &   22.181   &    0.010   &       r1   \\ 
  4277   &  62.442072   & $-$56.110460   &    1.099   &    0.074   &   0.0340   &   23.892   &    0.047   &       r1   \\ 
  4280   &  62.440208   & $-$56.113842   &    1.271   &    0.022   &   0.0266   &   21.119   &    0.011   &       r1   \\ 
  4320   &  62.436274   & $-$56.121193   &    1.004   &    0.039   &   0.0266   &   23.086   &    0.023   &       r1   \\ 
  4373   &  62.449126   & $-$56.102981   &    0.885   &    0.037   &   0.0725   &   23.182   &    0.022   &       r1   \\ 
  4436   &  62.453844   & $-$56.097291   &    0.919   &    0.026   &   0.0681   &   22.667   &    0.013   &       r1   \\ 
  4444   &  62.447418   & $-$56.107243   &    1.044   &    0.029   &   0.0274   &   22.096   &    0.015   &       r1   \\ 
  4526   &  62.444006   & $-$56.113942   &    1.260   &    0.030   &   0.0525   &   22.102   &    0.016   &       r1   \\ 
  4630   &  62.444749   & $-$56.114977   &    1.205   &    0.035   &   0.0237   &   22.418   &    0.020   &       r1   \\ 
  4658   &  62.449950   & $-$56.107424   &    1.203   &    0.043   &   0.0244   &   22.930   &    0.025   &       r1   \\ 
  4704   &  62.451416   & $-$56.106083   &    1.098   &    0.061   &   0.0163   &   23.611   &    0.040   &       r1   \\ 
  4742   &  62.460552   & $-$56.093062   &    1.112   &    0.021   &   0.0340   &   21.993   &    0.009   &       r1   \\ 
  4760   &  62.450792   & $-$56.108034   &    1.379   &    0.078   &   0.0858   &   23.664   &    0.045   &       r1   \\ 
  4790   &  62.454256   & $-$56.103414   &    0.909   &    0.026   &   0.0311   &   21.698   &    0.013   &       r1   \\ 
  4794   &  62.453328   & $-$56.104755   &    1.171   &    0.036   &   0.0266   &   22.670   &    0.020   &       r1   \\ 
  4806   &  62.452688   & $-$56.106034   &    1.164   &    0.049   &   0.0340   &   23.206   &    0.030   &       r1   \\ 
  4813   &  62.453102   & $-$56.105470   &    1.167   &    0.053   &   0.0770   &   23.371   &    0.032   &       r1   \\ 
  4888   &  62.460404   & $-$56.095814   &    1.182   &    0.062   &   0.0585   &   23.881   &    0.036   &       r1   \\ 
  4948   &  62.442253   & $-$56.124977   &    1.024   &    0.029   &   0.0503   &   22.215   &    0.016   &       r1   \\ 
  5095   &  62.449093   & $-$56.117356   &    0.884   &    0.048   &   0.0451   &   22.977   &    0.032   &       r1   \\ 
  5239   &  62.450679   & $-$56.117956   &    1.045   &    0.046   &   0.0400   &   22.773   &    0.029   &       r1   \\ 
  5271   &  62.448478   & $-$56.121895   &    1.364   &    0.056   &   0.1347   &   23.041   &    0.032   &       r1   \\ 
  5280   &  62.450171   & $-$56.119649   &    1.188   &    0.056   &   0.0466   &   23.132   &    0.036   &       r1   \\ 
  5290   &  62.457697   & $-$56.108462   &    0.974   &    0.029   &   0.0385   &   21.967   &    0.016   &       r1   \\ 
  5313   &  62.449563   & $-$56.121180   &    1.400   &    0.096   &   0.0266   &   23.858   &    0.061   &       r1   \\ 
  5389   &  62.449014   & $-$56.123932   &    0.882   &    0.021   &   0.0326   &   20.481   &    0.010   &       r1   \\ 
  5393   &  62.455059   & $-$56.114635   &    1.214   &    0.045   &   0.0133   &   22.601   &    0.028   &       r1   \\ 
  5446   &  62.463440   & $-$56.103252   &    1.140   &    0.035   &   0.0170   &   22.725   &    0.020   &       r1   \\ 
  5475   &  62.460226   & $-$56.108876   &    1.226   &    0.029   &   0.0969   &   21.923   &    0.016   &       r1   \\ 
  5604   &  62.450771   & $-$56.125846   &    1.230   &    0.029   &   0.0422   &   22.006   &    0.016   &       r1   \\ 
  5708   &  62.463336   & $-$56.108916   &    0.905   &    0.047   &   0.0215   &   22.901   &    0.031   &       r1   \\ 
  5754   &  62.463867   & $-$56.109087   &    1.221   &    0.052   &   0.1199   &   22.956   &    0.032   &       r1   \\ 
  5768   &  62.452947   & $-$56.125993   &    0.952   &    0.034   &   0.0385   &   21.897   &    0.016   &       r1   \\ 
  5837   &  62.456743   & $-$56.121757   &    0.887   &    0.026   &   0.0281   &   21.300   &    0.014   &       r1   \\ 
  5865   &  62.469218   & $-$56.103303   &    1.208   &    0.030   &   0.0681   &   22.459   &    0.016   &       r1   \\ 
  5963   &  62.455278   & $-$56.127144   &    0.887   &    0.038   &   0.0141   &   22.545   &    0.024   &       r1   \\ 
  6042   &  62.466243   & $-$56.111291   &    1.341   &    0.120   &   0.0067   &   23.842   &    0.077   &       r1   \\ 
  6050   &  62.469475   & $-$56.107693   &    1.184   &    0.022   &   0.0296   &   20.902   &    0.011   &       r1   \\ 
  6052   &  62.472359   & $-$56.103277   &    1.100   &    0.028   &   0.0237   &   22.218   &    0.015   &       r1   \\ 
  6055   &  62.474468   & $-$56.100142   &    0.918   &    0.033   &   0.0237   &   23.026   &    0.019   &       r1   \\ 
  6058   &  62.458180   & $-$56.124987   &    1.226   &    0.061   &   0.0340   &   23.158   &    0.038   &       r1   \\ 
  6179   &  62.474633   & $-$56.101859   &    1.084   &    0.027   &   0.0237   &   22.296   &    0.014   &       r1   \\ 
  6279   &  62.473238   & $-$56.107092   &    1.177   &    0.051   &   0.0200   &   23.148   &    0.032   &       r1   \\ 
  6429   &  62.474914   & $-$56.107999   &    1.263   &    0.047   &   0.0385   &   22.968   &    0.029   &       r1   \\ 
  6457   &  62.475410   & $-$56.108026   &    0.856   &    0.031   &   0.0296   &   22.152   &    0.018   &       r1   \\ 
  6597   &  62.484653   & $-$56.097083   &    1.222   &    0.031   &   0.0525   &   23.020   &    0.015   &       r1   \\ 
  6622   &  62.480808   & $-$56.099170   &    1.087   &    0.023   &   0.0266   &   21.884   &    0.010   &       r1   \\ 
  6842   &  62.480776   & $-$56.108464   &    0.986   &    0.038   &   0.0170   &   22.715   &    0.023   &       r1   \\ 
  6872   &  62.490483   & $-$56.094283   &    0.853   &    0.033   &   0.0599   &   23.327   &    0.016   &       r1   \\ 
  6997   &  62.483392   & $-$56.107559   &    1.183   &    0.042   &   0.0244   &   23.067   &    0.025   &       r1   \\ 
  7017   &  62.471658   & $-$56.125870   &    0.971   &    0.044   &   0.0215   &   22.527   &    0.028   &       r1   \\ 
  7081   &  62.484164   & $-$56.108342   &    0.910   &    0.030   &   0.0340   &   22.408   &    0.017   &       r1   \\ 
  7213   &  62.476729   & $-$56.122510   &    0.841   &    0.054   &   0.0503   &   23.030   &    0.036   &       r1   \\ 
  7455   &  62.494309   & $-$56.100625   &    0.945   &    0.036   &   0.0296   &   22.844   &    0.018   &       r1   \\ 
  7509   &  62.494724   & $-$56.101300   &    0.892   &    0.026   &   0.0363   &   22.653   &    0.013   &       r1   \\ 
\enddata
\label{tab:dat}
\end{deluxetable}

\end{document}